\documentclass[final,a4paper,11pt]{article}
\pdfoutput=1

\usepackage{jheppub} 

\usepackage[T1]{fontenc}
\usepackage{natbib}
\usepackage[utf8]{inputenc}

\usepackage{hyperref}
\usepackage{bbold}

\usepackage{graphicx,epsf}
\usepackage{amsmath,amsfonts,amssymb,amsbsy}

\usepackage{hyperref}
\usepackage{xcolor}
\hypersetup{
    colorlinks=true,
    allcolors={blue!60!black}
}

\usepackage{multirow}
\usepackage{stackrel}
\usepackage{tikz}
\usetikzlibrary{calc}

\usepackage{ifdraft}
\usepackage[obeyFinal]{easy-todo}

\usepackage{placeins}
\usepackage{slashed} 

\usepackage{booktabs}
\usepackage{adjustbox}

\usepackage{cleveref}

\DeclareMathOperator{\Tr}{Tr}

\newcommand{\eps}{\epsilon}

\def\NC{{N_c}}
\def\NF{{N_f}}
\def\CA{\mathcal{A}}

\def\Mathematica{{\sc Mathematica}}

\def\trFive{{\rm tr}_5}

\begin{document}

\title{
    Analytic Form of the Planar Two-Loop Five-Parton Scattering Amplitudes in QCD
}
\preprint{CP3-19-13, FR-PHENO-2019-002, IPhT-19/024}

\author[a]{S.~Abreu,}
\author[b]{J.~Dormans,}
\author[b,c]{F.~Febres~Cordero,}
\author[b]{H.~Ita,}
\author[d]{B.~Page,}
\author[b]{and V.~Sotnikov}

\affiliation[a]{Center for Cosmology, Particle Physics and
Phenomenology (CP3), Universit\'{e} Catholique de Louvain, 1348
Louvain-La-Neuve, Belgium}
\affiliation[b]{Physikalisches Institut,
Albert-Ludwigs-Universit\"at Freiburg, Hermann-Herder-Str.~3, 
D-79104 Freiburg, Germany}
\affiliation[c]{Physics Department, Florida State University,
  77 Chieftan Way, Tallahassee, FL 32306, U.S.A.}
\affiliation[d]{Institut de Physique Th\'eorique, CEA, CNRS, 
Universit\'e Paris-Saclay, F-91191 Gif-sur-Yvette cedex, France}

\abstract{
We present the analytic form of all leading-color two-loop five-parton
helicity amplitudes in QCD. The results are analytically reconstructed
from exact numerical evaluations over finite fields.
Combining a judicious choice of variables with 
a new approach to the treatment of particle states in $D$ dimensions
for the numerical evaluation of amplitudes, we
obtain the analytic expressions with a modest computational effort.
Their systematic simplification using multivariate partial-fraction
decomposition leads to a particularly compact form.
Our results provide all two-loop
amplitudes required for the calculation of next-to-next-to-leading order
QCD corrections to the production of three jets at hadron colliders in
the leading-color approximation.
}

\maketitle

\ifoptionfinal{}{\listoftodos}


\section{Introduction}\label{sec:intro}

Scattering amplitudes in QCD provide the basic building blocks for hadron
collider phenomenology. Integrated over phase space, they yield theoretical
predictions that can be compared to experimental measurements like those
performed at the CERN Large Hadron Collider (LHC). These theoretical
predictions can be systematically improved through inclusion of higher-order
QCD corrections, which require higher-loop scattering amplitudes.  The analytic
computation of these multi-loop amplitudes is a non-trivial task.

Experience at one loop has shown that the direct numerical computation of
scattering amplitudes can avoid difficulties encountered in analytic
approaches. However, while numerical evaluation is sufficient for
phenomenological applications, compact analytic results are still very useful.  
Indeed, the numerical evaluation of analytic expressions is
generally straightforward to set up and oftentimes more stable and efficient
than the purely numerical approach.  Furthermore, explicit formulae allow for a
detailed analysis of the analytic properties of amplitudes in order to learn
about  higher-order perturbation theory.

In this article we present the analytic expressions of the two-loop five-parton
QCD amplitudes in the leading-color approximation.  Recently, significant
progress has been made regarding the computation of two-loop multi-particle
amplitudes.  In the case of five-point QCD amplitudes, the first one to be studied
was the leading-color five-gluon amplitude with all helicities positive, initially
evaluated numerically~\cite{Badger:2013gxa} and  
afterwards presented analytically~\cite{Gehrmann:2015bfy,Dunbar:2016aux}.
Building on this result, the all-plus two-loop six- and seven-gluon amplitudes were
obtained~\cite{Dunbar:2016gjb,Dunbar:2017nfy}.  By now all leading-color
two-loop five-parton amplitudes (i.e., with external gluons and/or
massless quarks) have been computed
numerically~\cite{Badger:2017jhb, Abreu:2017hqn, Badger:2018gip,
Abreu:2018jgq}. Very recently, numerical algorithms were combined with
functional reconstruction techniques~\cite{Peraro:2016wsq} resulting in
analytic expressions for the planar two-loop five-gluon single-minus helicity
amplitude~\cite{Badger:2018enw} and for all two-loop five-gluon helicity
amplitudes~\cite{Abreu:2018zmy}. These calculations rely on the availability
of the planar two-loop five-point master
integrals, which have been given in refs.~\cite{Papadopoulos:2015jft,Gehrmann:2018yef}. 
Progress with full-color five-gluon amplitudes~\cite{Badger:2015lda} is 
gaining momentum with recent results towards the computation of non-planar master
integrals~\cite{Abreu:2018aqd,Chicherin:2018old}.
In more conventional approaches, integration-by-parts relations were
obtained~\cite{Boels:2018nrr,Chawdhry:2018awn} which can be used to compute the
same type of two-loop five-point amplitudes.

Here we apply a numerical
variant~\cite{Ossola:2006us,Ellis:2007br,Giele:2008ve,Berger:2008sj} of the
unitarity method~\cite{Bern:1994zx,Bern:1994cg,Bern:1997sc,Britto:2004nc} which
has recently been extended to two
loops~\cite{Ita:2015tya,Abreu:2017idw,Abreu:2017xsl}. Furthermore, we take
advantage of computations with exact kinematics through the usage of
finite-field arithmetic~\cite{vonManteuffel:2014ixa,Peraro:2016wsq} in order to
functionally reconstruct~\cite{Peraro:2016wsq} the multivariate rational
coefficients of a basis of special functions~\cite{Gehrmann:2018yef}. We have
already shown that this strategy can be employed to compute amplitudes of
relevance to LHC phenomenology, by providing the analytic expressions of all
planar two-loop five-gluon amplitudes~\cite{Abreu:2018zmy}. In this article we
further improve on the latter work by producing compact analytic results for
all leading-color two-loop five-parton amplitudes in QCD.  This includes the
five-parton processes with five gluons, two quarks and three gluons, and four
quarks and one gluon, with zero, one and two light-quark loops.  These are all
the two-loop amplitudes required for the computation of the leading-color
next-to-next-to-leading order (NNLO) QCD corrections to three-jet production at
hadron colliders.

A number of new developments allows to obtain these results.
First, we set up an efficient method to determine the dependence on the
dimension $D_s$ associated to the particle states circulating in the loop.  We
extend the approach of~\cite{Anger:2018ove} and remove a bottleneck of
dimensional reconstruction~\cite{Giele:2008ve,Ellis:2008ir, Boughezal:2011br}
in fermion amplitudes~\cite{Abreu:2018jgq} by analytically precomputing part of
the dependence on the dimensional regulator.

Second, we modify the multivariate reconstruction algorithm that we employed
in ref.~\cite{Abreu:2018zmy} in order to use an ansatz in terms of Mandelstam variables
instead of twistor variables~\cite{Hodges:2009hk}. This is achieved by
exploiting the analytic structure of amplitudes with five external massless
partons in order to obtain target functions which are rational functions of
Mandelstam variables. This has a dramatic impact in reducing the degree of the
multivariate polynomials that we reconstruct, allowing us to obtain all
analytic expressions with a modest computational effort.

Finally, in order to employ a single finite field for rational
reconstruction, we perform a systematic analysis of the analytic structure of
the reconstructed functions which takes advantage of two major simplification
procedures. As a by-product, the procedure results in rather compact expressions
for all helicity amplitudes, which we provide as ancillary files. 
We first analyze the dimension of the function
space spanned by all the pentagon-function coefficients on each amplitude. It
turns out that this dimension is an order of magnitude smaller (on average)
than the number of pentagon functions. Then we simplify each member of the
basis of this space with a multivariate partial-fraction decomposition.  All
taken into account, the final expressions we present are reduced in byte size
by up to two orders of magnitude, with final results that can easily be handled
for future analytic and numerical evaluations.

This article is organized as follows. Section \ref{sec:setup} is devoted to the
description of the numerical calculation of the multi-parton amplitudes. There
we precisely define the objects we compute, including the two-loop helicity
amplitudes and the finite remainder functions in dimensional regularization.
In section \ref{sec:Ds} our method of handling the $D_s$~dependence is
presented, and section \ref{sec:reconstruction} describes the analytic
reconstruction and the simplification of the results by means of multivariate
partial fractioning. We discuss the analytic results in section
\ref{sec:results} before concluding in section \ref{sec:conclusions}.
Additional information related to the infrared structure of the amplitudes,
Feynman rules and a rationalization of the momentum-space variables is
presented in three appendices.

\section{Calculation of Multi-Parton Planar Amplitudes}\label{sec:setup}

\subsection{Multi-Parton Helicity Amplitudes}\label{sec:amps}

In this work we compute the
five-parton two-loop QCD helicity amplitudes
in the leading-color approximation. More precisely, we keep 
the leading terms in the formal limit of a large number of
colors $\NC$, and scale the number of light flavors $\NF$ 
while keeping the ratio $\NF/\NC$ fixed. 

We evaluate amplitudes in dimensional regularization.  Care must be
taken when applying dimensional regularization in a numerical approach
with external fermions \cite{Abreu:2018jgq}. In such an approach
amplitudes are computed with integer dimensional particle
representations, while dimensionally-regulated amplitudes are
analytically continued to non-integer dimensions $D=4-2\epsilon$.  We
will follow the same procedure described in detail in
ref.~\cite{Abreu:2018jgq}, which is related to the idea of using
suitable projection operators \cite{Glover:2004si} allowing one to work
with Lorentz-invariant objects in the main computational steps.  A 
scattering amplitude $M$ can thus be written
as
\begin{equation}
  M = \sum_n  v_n M_n\, ,
\label{eq:tensorDecomposition}
\end{equation}
where the $M_n$ are Lorentz scalars and the $v_n$ provide a basis
for the $(D_s-4)$-dimensional spinor structures. 
The basis $\{v_n\}$ is
process dependent, and the loop order at which 
a given spinor structure begins to contribute
depends on the regularization
scheme.  For concreteness, we work in the 't~Hooft-Veltman (HV) scheme 
and use the bases of $\{v_n\}$ considered in ref.~\cite{Abreu:2018jgq}.  
For NNLO phenomenology, in cases where
two-loop virtual corrections are interfered with a tree amplitude, it is
sufficient to compute $M_0$ through
\begin{subequations}
  \label{eq:defAmpsTens}
  \begin{align}
    A(q,\bar q,g,\ldots,g)
    &\equiv \delta_\lambda^\kappa
    \left(M(q,\bar q,g,\ldots,g)\right)_\kappa^\lambda
    \,,\\
    A(q,\bar q,Q,\bar Q,g,\ldots,g)
    &\equiv \delta_{\lambda_1}^{\kappa_1}
    \delta_{\lambda_2}^{\kappa_2}
    \left(M(q,\bar q,Q,\bar Q,g,\ldots,g)
    \right)_{\kappa_1\kappa_2}^{\lambda_1\lambda_2}\,,
    \label{eq:defAmpsTensb}
  \end{align}
\end{subequations}
where on the right-hand-side of the equations we compute the traces to
project onto the structures $v_0$ in eq.~\eqref{eq:tensorDecomposition}.
This means that the relevant contributions are those in which
the open $(D_s-4)$-dimensional spinor indices are traced over on
each quark line, see figure~\ref{fig_traces}.

While the amplitudes in eq.~\eqref{eq:defAmpsTens} are easily
defined and evaluated analytically, it is important to find an
efficient way to compute them in a numerical setup. We discuss our
approach in section~\ref{sec:Ds} where we give details of our
implementation for the numerical evaluation of amplitudes with fermions.
This definition of helicity
amplitudes with quarks is consistent with that of ref.~\cite{Glover:2004si} and 
the prescription given in ref.~\cite{Anger:2018ove}. 

When considering amplitudes with two identical quark lines, contributions where
index contractions lead to a single trace as in figure~\ref{fig_SingleTrace} 
should also be considered. Nevertheless, at the level of the
finite remainder, amplitudes with identical quarks can be obtained by
antisymmetrizing distinct-flavor
expressions~\cite{DeFreitas:2004kmi,Abreu:2018jgq} obtained from
eq.~\eqref{eq:defAmpsTensb}. Thus, our results are sufficient for 
NNLO QCD phenomenological studies of processes involving 
identical quarks.

\begin{figure}[]
  \begin{center}
    \begin{tikzpicture}[scale=.9]
    \node at (0,0){\includegraphics[scale=0.5]
    {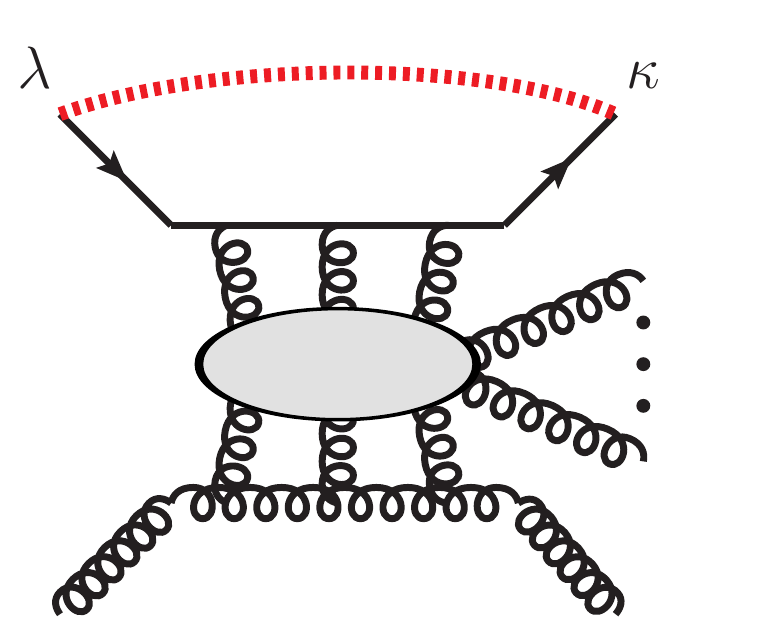}};
    \node at (7,-.1){\includegraphics[scale=0.5]
    {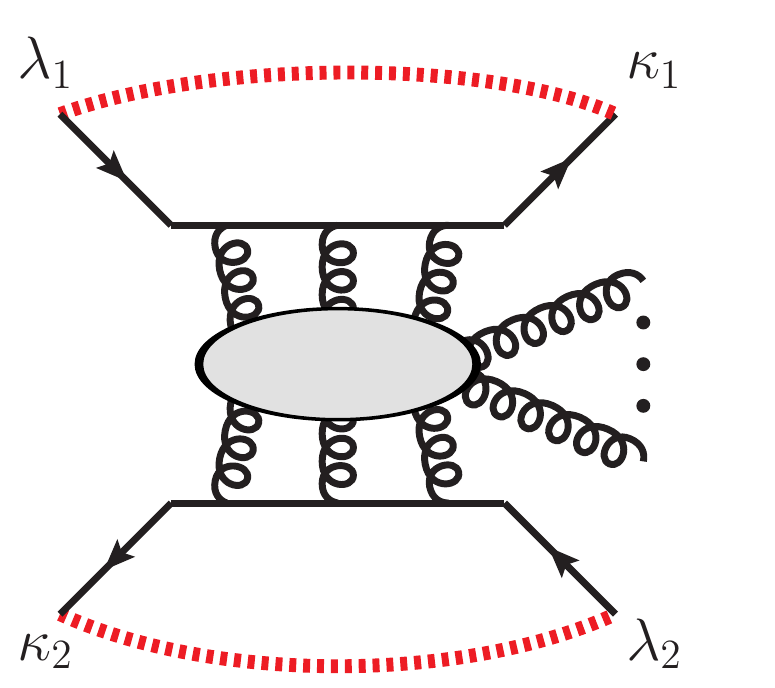}};
\end{tikzpicture}
\end{center} 
\caption{Contraction of the open $(D_s-4)$-dimensional spinor indices for amplitudes
with external quarks: each quark line closes upon itself.
Indices connected by red dashed lines are traced over.}
\label{fig_traces}
\end{figure}
\begin{figure}[]
  \begin{center}
  	\includegraphics[scale=0.5]
    {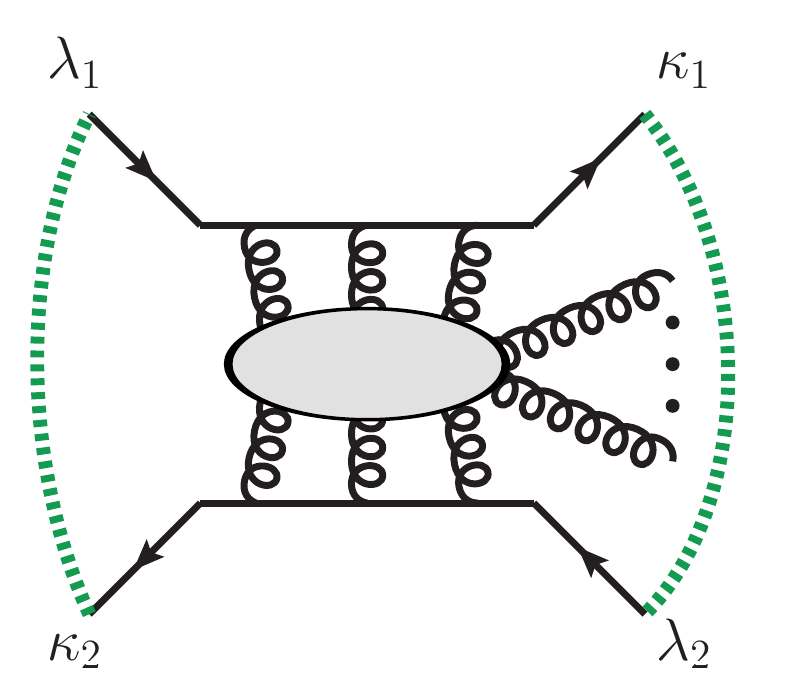}
\end{center} 
\caption{Alternative contraction of $(D_s-4)$-dimensional spinor indices leading to a
single trace. Indices connected by green dashed lines are traced
over.}
\label{fig_SingleTrace}
\end{figure}

The gluon and fermion amplitudes \eqref{eq:defAmpsTens}
can be decomposed 
in terms of color structures. 
We denote the fundamental generators of the
$SU(\NC)$-group by $(T^a)^{\;\bar{\jmath}}_{i}$, where the
adjoint index $a$ runs over $\NC^2-1$ values and the 
(anti-) fundamental indices $i$  ($\bar \jmath$)
run over $\NC$ values. The fundamental generators are normalized
as $ \Tr(T^a T^b) = \delta^{ab}$. One can then consider the color decomposition of
each process as
\begin{align}
  \begin{split}
  \label{eq:ColorDecG}
  A(1_g, 2_g, 3_g, 4_g, 5_g) \big\vert_{\textrm{leading color}} = &
  \sum_{\sigma\in S_5/Z_5} \Tr\left(
  T^{a_{\sigma(1)}} T^{a_{\sigma(2)}} 
  T^{a_{\sigma(3)}} T^{a_{\sigma(4)}} T^{a_{\sigma(5)}} \right)\\
  &\times \CA({\sigma(1)}_g, {\sigma(2)}_g, {\sigma(3)}_g, {\sigma(4)}_g, {\sigma(5)}_g)\,, 
  \end{split} \\
  \begin{split}
  \label{eq:ColorDec2Q}
  A(1_q, 2_{\bar{q}}, 3_g, 4_g, 5_g) \big\vert_{\textrm{leading color}} 
    = & \sum_{\sigma\in S_3} 
  \left( T^{a_{\sigma(3)}} T^{a_{\sigma(4)}} T^{a_{\sigma(5)}} \right)^{\;\bar{\imath}_2}_{i_1} \\
  & \times\CA(1_q,2_{\bar{q}},\sigma(3)_g,\sigma(4)_g,\sigma(5)_g)\,,
  \end{split}\\
  \begin{split}
  \label{eq:ColorDec4Q}
  A(1_q, 2_{\bar{q}}, 3_Q, 4_{\bar{Q}}, 5_g)
  \big\vert_{\textrm{leading color}} 
  = & 
  \,(T^{a_5})^{\;\bar{\imath}_{2}}_{i_{3}} \delta^{\;\bar{\imath}_{4}}_{i_{1}}\;
  \CA(1_{q}, 2_{\bar{q}}, 5_g, 3_{Q}, 4_{\bar{Q}}) \,\,+  \\
  & \,(T^{a_5})^{\;\bar{\imath}_{4}}_{i_{1}} \delta^{\;\bar{\imath}_{2}}_{i_{3}}\;
\CA(1_{q},2_{\bar{q}},3_{Q},4_{\bar{Q}},5_g) \,,
\end{split}
\end{align}
where $S_n$ denotes all permutations of $n$ indices and 
$S_n/Z_n$ all non-cyclic permutations of $n$ indices. 
We write the particle type explicitly as a subscript, and all
remaining properties of each particle (momentum, helicity, 
etc.) are implicit. 

The $\CA$ in \cref{eq:ColorDecG,eq:ColorDec2Q,eq:ColorDec4Q} are
called partial amplitudes. For each of the amplitudes considered, all
partial amplitudes can be related to the others by exchanging external
legs, so only one partial amplitude is independent. These are expanded in a perturbative
expansion,
\begin{equation}
    \label{eq:partials} 
    \CA
    = g^{3}_0 \left(
        \CA^{(0)}
      + \frac{\alpha_0}{4\pi}\NC \CA^{(1)}
      + \left(\frac{\alpha_0}{4\pi}\right)^2\NC^2  \CA^{(2)} 
      + \mathcal{O}(\alpha_0^3)
      \right),
\end{equation}
where $\alpha_0=g_0^2/(4\pi)$ is the bare QCD coupling and 
$\CA^{(k)}$ denotes a $k$-loop partial amplitude. Each 
$\CA^{(k)}$ can be further expanded as a series in powers of 
$N_f/N_c$,
\begin{align}
  \label{eq:nfdecomposition} 
  \begin{split}
  \CA^{(1)} &= \CA^{(1)[N_f^0]} + 
  \frac{\NF}{\NC}\CA^{(1)[N_f^1]}\,, \\
  \CA^{(2)} &= \CA^{(2)[N_f^0]} +
  \frac{\NF}{\NC}\CA^{(2)[N_f^1]} +
  \left(\frac{\NF}{\NC}\right)^2\CA^{(2)[N_f^2]}\,.
  \end{split} 
\end{align}
We compute the coefficients $\CA^{(k)[N_f^l]}$, with $0\leq l\leq
k\leq2$, where only planar diagrams contribute as we work in the
leading-color approximation. In figs.~\ref{fig_parents5g},
\ref{fig_parents2q3g} and \ref{fig_parents4q1g} we give representative
diagrams for each of these contributions.

\begin{figure}[]
  \begin{center}
    \begin{tikzpicture}[scale=.9]
    \node at (0,0){\includegraphics[scale=0.5]
    {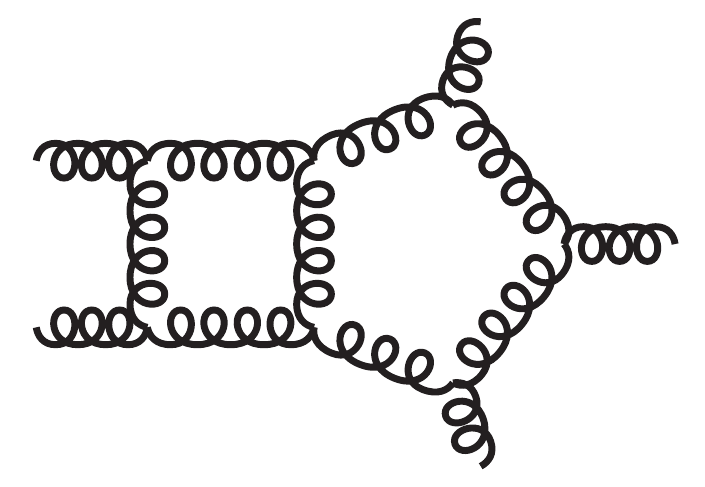}};
    \node at (5,0){\includegraphics[scale=0.5]
    {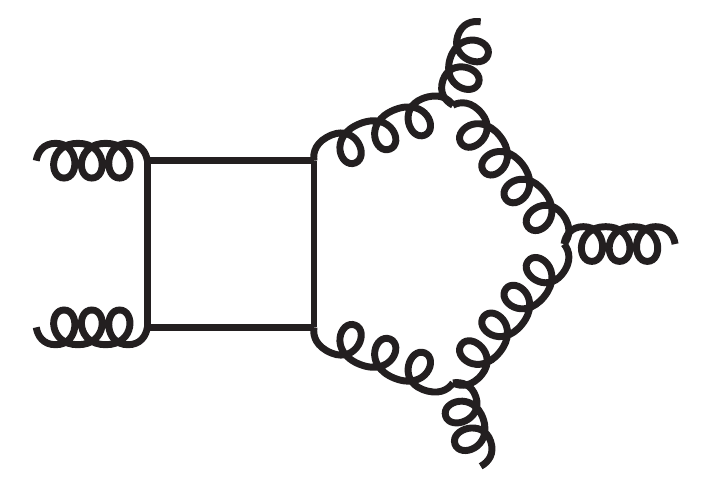}};
    \node at (10,0){\includegraphics[scale=0.5]
    {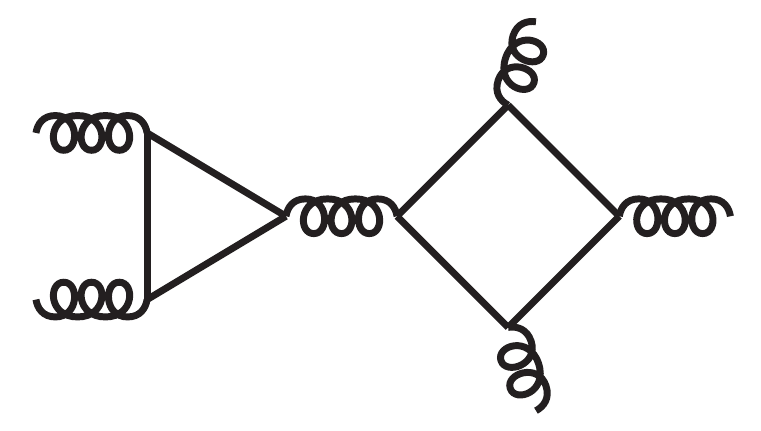}};
\end{tikzpicture}
\end{center} 
\caption{Representative Feynman diagrams for leading-color
$\CA^{(2)}(g,g,g,g,g)$ amplitudes, contributing at order
$N_f^0$, $N_f^1$ and $N_f^2$.}
\label{fig_parents5g}
\end{figure}

\begin{figure}[]
  \begin{center}
    \begin{tikzpicture}[scale=.9]
    \node at (0,0){\includegraphics[scale=0.5]
    {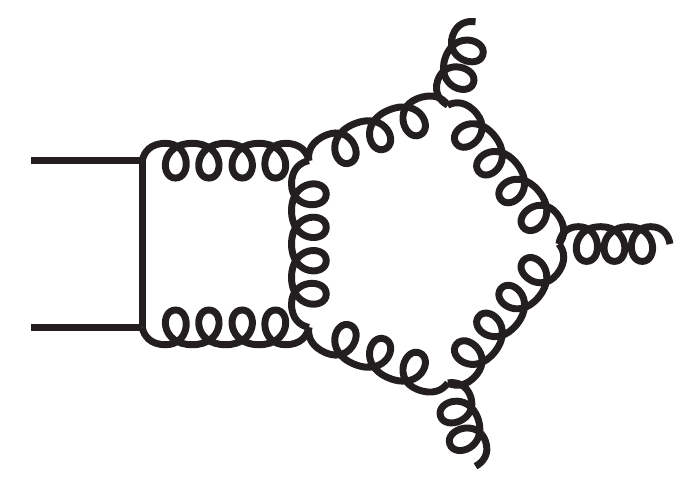}};
    \node at (5,0){\includegraphics[scale=0.5]
    {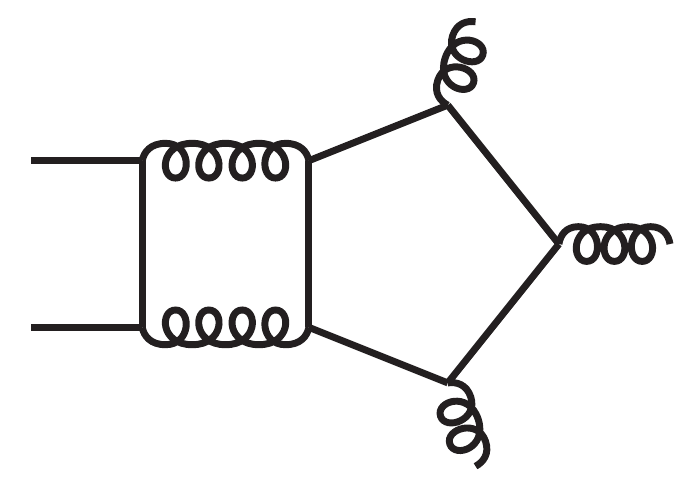}};
    \node at (10,0){\includegraphics[scale=0.5]
    {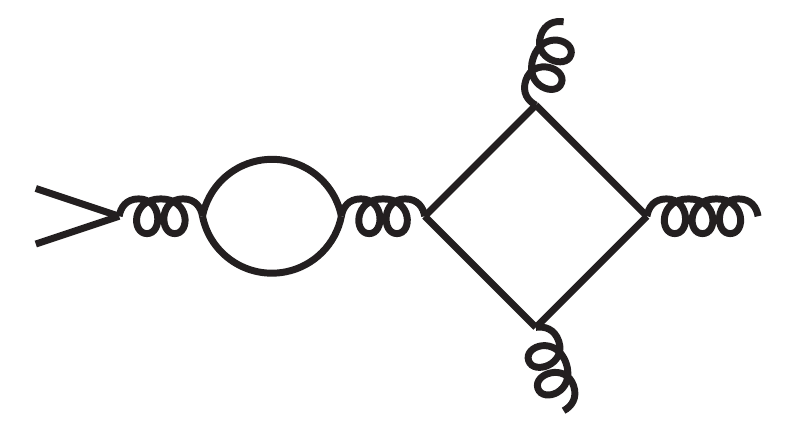}};
\end{tikzpicture}
\end{center} 
\caption{Representative Feynman diagrams for leading-color
$\CA^{(2)}(q,\bar q,g,g,g)$ amplitudes, 
contributing at order
 $N_f^0$, $N_f^1$ and $N_f^2$.}
\label{fig_parents2q3g}
\end{figure}

\begin{figure}[]
  \begin{center}
    \begin{tikzpicture}[scale=.9]
    \node at (0,0){\includegraphics[scale=0.5]
    {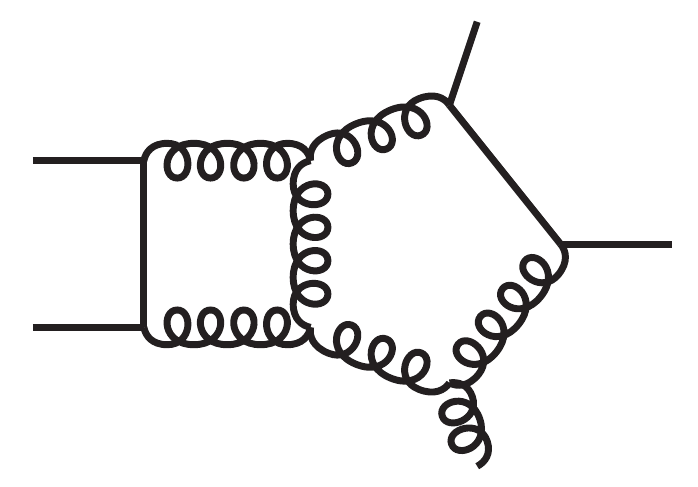}};
    \node at (5,.4){\includegraphics[scale=0.5]
    {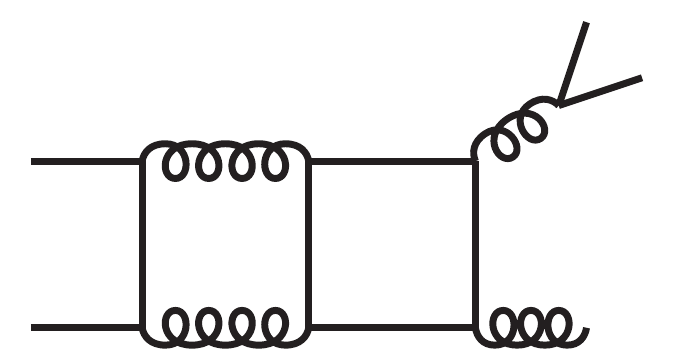}};
    \node at (10,.4){\includegraphics[scale=0.5]
    {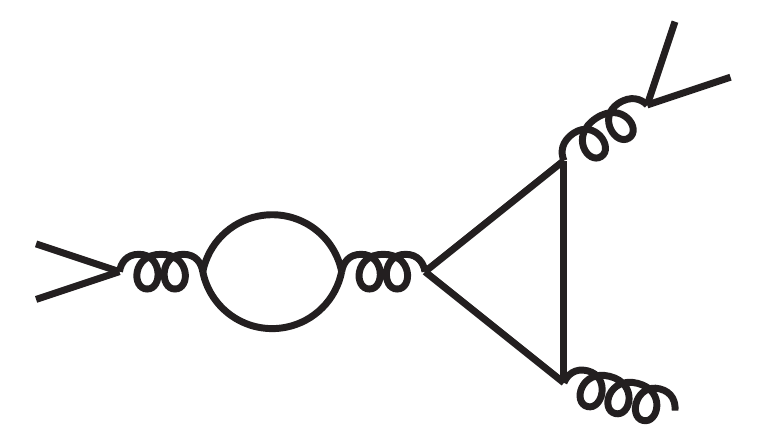}};
\end{tikzpicture}
\end{center} 
\caption{Representative Feynman diagrams for leading-color
$\CA^{(2)}(q,\bar q,Q,\bar Q,g)$ amplitudes, 
contributing at order
$N_f^0$, $N_f^1$ and $N_f^2$.}
\label{fig_parents4q1g}
\end{figure}

\subsection{Finite Remainder}\label{sec:remainders}

The bare scattering amplitudes defined in 
eq.~\eqref{eq:nfdecomposition} 
have divergences of ultraviolet and 
infrared origin. Both can be predicted from lower-loop
amplitudes. It is convenient to remove this redundant information
and define a finite remainder that contains the genuine two-loop
information. There is no unique way to define the remainder, so we now 
discuss our conventions, with more details given in 
appendix~\ref{app:remainderDetails}.

The renormalized amplitudes can be obtained from their bare counterparts
by replacing in eq.~\eqref{eq:partials} the bare QCD coupling $\alpha_0$
by the renormalized coupling $\alpha_s$ in $D=4-2\epsilon$ dimensions. 
The two couplings are related by
\begin{equation}\label{eq:renormCoupling}
    \alpha_0\mu_0^{2\epsilon}S_{\epsilon}
  =\alpha_s\mu^{2\epsilon}\left(
  1-\frac{\beta_0}{\epsilon}\frac{\alpha_s}{4\pi}
  +\left(\frac{\beta_0^2}{\epsilon^2}-\frac{\beta_1}{\epsilon}\right)
  \left(\frac{\alpha_s}{4\pi}\right)^2+\mathcal{O}
  \left(\alpha_s^3\right)\right),
\end{equation}
which we can use to define the perturbative expansion of the renormalized
amplitude,
\begin{equation}\label{eq:renormAmp}
  \mathcal{A}_R = S_\epsilon^{-\frac{3}{2}}
  g_s^3\left(
  \mathcal{A}_R^{(0)}
  +\frac{\alpha_s}{4\pi}\NC\,\mathcal{A}_R^{(1)}
  +\left(\frac{\alpha_s}{4\pi}\right)^2\NC^2\mathcal{A}_R^{(2)}
  +\mathcal{O}(\alpha_s^3)
  \right),
\end{equation}
where $S_\epsilon=(4\pi)^{\epsilon}e^{-\epsilon\gamma_E}\,$ and 
$\alpha_s=g_s^2/(4\pi)$. The $\beta_i$ are the coefficients in the
perturbative expansion of the QCD $\beta$-function, which we give explicitly
in appendix \ref{app:remainderDetails}. 
Here, $\mu_0^2$ is the scale introduced in dimensional regularization to keep the
coupling dimensionless in the QCD Lagrangian, and $\mu^2$ is the
renormalization scale. In the following, we set $\mu^2=\mu_0^2=1$ (with arbitrary dimensions).

The renormalized amplitudes $\CA^{(k)}_R$ have only infrared divergences,
which can be determined from lower-loop functions and well known universal 
factors \cite{Catani:1998bh,Sterman:2002qn,Becher:2009cu,Gardi:2009qi}. More
precisely, we have
\begin{align}\begin{split}\label{eq:catani}
    \CA_R^{(1)}&={\bf I}^{(1)}_{[n]}(\epsilon)
    \CA_R^{(0)}+\mathcal{O}
    (\epsilon^0)\,,\\
    \CA_R^{(2)}&={\bf I}^{(2)}_{[n]}(\epsilon)\CA_R^{(0)}+{\bf I}^{(1)}_{[n]}(\epsilon)
    \CA_R^{(1)}+\mathcal{O}(\epsilon^0)\,.
\end{split}\end{align}
The ${\bf I}^{(k)}_{[n]}$ are operators in color space that become diagonal
in the leading-color approximation. They contain some
process-specific components, and explicit expressions are given 
in appendix~\ref{app:remainderDetails}.

Using \cref{eq:renormCoupling,eq:renormAmp,eq:catani}, we can predict
the poles of the two-loop amplitudes we wish to compute. Alternatively, we
can use them to define a finite remainder $\mathcal{R}^{(2)}$, according to
\begin{equation}\label{eq:remaindeDef}
  \mathcal{R}^{(2)}=\mathcal{A}_R^{(2)}
  -{\bf I}_{[n]}^{(1)}\mathcal{A}_R^{(1)}
  -{\bf I}_{[n]}^{(2)}\mathcal{A}_R^{(0)}
  +\mathcal{O}(\epsilon)\,.
\end{equation}
In this expression, we extend the expansion of 
${\bf I}_{[n]}^{(1)}\mathcal{A}_R^{(1)}$ and
${\bf I}_{[n]}^{(2)}\mathcal{A}_R^{(0)}$ to also include
terms of order $\eps^0$. This subtracts non-trivial contributions from
the finite term of $\mathcal{A}_R^{(2)}$ that are related to 
the lower-loop amplitudes. In this paper, we directly compute 
analytic expressions for the remainders $\mathcal{R}^{(2)}$, from which
one can then recover the full two-loop bare amplitude $\CA^{(2)}$,
see eq.~\eqref{eq:ampFromRem} for the explicit relation.

\subsection{Numerical Unitarity}\label{sec:unitarity}

The first step in the analytic reconstruction of the remainders defined
in eq.~\eqref{eq:remaindeDef} is to evaluate the bare amplitudes
numerically. To this end, we employ numerical
unitarity~\cite{Ita:2015tya,Abreu:2017xsl,Abreu:2017idw}
in order to evaluate
the coefficients of a decomposition of the amplitude into a
linear combination of master integrals.
First, we consider a decomposition of the integrand of the amplitude
$\mathcal{A}^{(2)}(\ell_l)$ in terms of master integrands
and surface terms (we use $\ell_l$ to denote the loop momenta). 
The former correspond to master integrals and the latter integrate to zero. 
Specifically, we have
\begin{equation}\label{eq:AL}
    \mathcal{A}^{(2)}(\ell_l)=\sum_{\Gamma\in\Delta}
    \sum_{i\in M_\Gamma\cup S_\Gamma} c_{\Gamma,i}
    \frac{m_{\Gamma,i}(\ell_l)}{\prod_{j\in
    P_\Gamma}\rho_j}\, ,
\end{equation}
where $\Delta$ is the set of all propagator structures $\Gamma$, 
$P_\Gamma$ the set of all inverse propagators $\rho_j$ in~$\Gamma$, 
and $M_\Gamma$ and $S_\Gamma$ denote the corresponding sets of 
master integrands and surface terms. 

To compute the decomposition of the amplitude in terms of master 
integrals we must evaluate the coefficients $c_{\Gamma,i}$, 
with $i\in M_\Gamma$. These are rational functions of spinor components
of the external momenta and the dimensional regulator $\epsilon$. We determine
them using the standard approach in numerical unitarity.
We first build a system of linear equations through 
sampling of on-shell values of the loop momenta, i.e., where 
$\ell_l\to\ell_l^\Gamma$ with $\ell_l^\Gamma$ such that 
$\rho_j(\ell_l^\Gamma)=0$ for $j\in P_\Gamma$. The leading contribution
to eq.~\eqref{eq:AL} in this limit factorizes into products of tree
amplitudes
\begin{equation}
\sum_{\rm states}\prod_{i\in T_\Gamma}
{\cal A}^{(0)}_i(\ell_l^\Gamma) =
\sum_{\substack{\Gamma' \ge \Gamma\,,\\ 
i\,\in\,M_{\Gamma'}\cup S_{\Gamma'}}} 
\frac{ c_{\Gamma',i}\,m_{\Gamma',i}(\ell_l^\Gamma)}{\prod_{j\in
(P_{\Gamma'}\setminus P_\Gamma) } \rho_j(\ell_l^\Gamma)}\,,
\label{eq:CE}
\end{equation}
where we label the set of tree amplitudes associated with the 
vertices in the diagram corresponding to $\Gamma$ by $T_\Gamma$,
and the sum on the right-hand side runs over the propagator structures
$\Gamma'$ such that $P_\Gamma\subseteq P_{\Gamma'}$. The sum on the 
left-hand side runs over the (scheme dependent) 
physical states of each internal line of $\Gamma$. At two loops
there are also subleading contributions in the limit 
$\ell_l\to\ell_l^\Gamma$ which can easily be dealt with even though
no factorization theorem is known \cite{Abreu:2017idw} (for a recent related
study on these contributions see~\cite{Baumeister:2019rmh}).
The coefficients $c_{\Gamma,i}$ can then be obtained at a given
phase-space point by solving the linear system in 
eq.~\eqref{eq:CE} numerically.
The numerical evaluations are performed using finite-field arithmetic, allowing
to efficiently obtain exact results for rational phase-space points,
circumventing problems of numerical instabilities. This is key for the task of
functional reconstruction.

\subsection{Pentagon-Function Decomposition}\label{sec:pentagon}

Once the coefficients $c_{\Gamma,i}$ in eq.~\eqref{eq:AL} are computed,
we obtain a decomposition of the amplitude into master integrals,
\begin{equation}
  \mathcal{A}^{(2)}=\sum_{\Gamma\in\Delta}
  \sum_{i\in M_\Gamma} c_{\Gamma,i}\,
  \mathcal{I}_{\Gamma,i}\,,
  \label{eq:Amaster}
\end{equation}
where the $\mathcal{I}_{\Gamma,i}$ are the master integrals,
\begin{equation}
  \mathcal{I}_{\Gamma,i}=\int d^D\ell_l
  \frac{m_{\Gamma,i}(\ell_l)}
  {\prod_{j\in P_\Gamma}\rho_j}\,.
\end{equation}
For planar five-parton amplitudes, a basis of master integrals has been 
computed~\cite{Papadopoulos:2015jft,Gehrmann:2018yef}.
The master integrals evaluate to linear combinations of so-called 
multiple polylogarithms (MPLs), which can be 
numerically evaluated using available programs 
(e.g.~\cite{Vollinga:2004sn}). Once numerical values for the coefficients 
$c_{\Gamma,i}$ have been  computed (using for example 
the approach described in the previous subsection) one can obtain numerical values
for the amplitudes.

The MPLs are a class of special functions with 
only logarithmic singularities that can be equipped with algebraic
structures that allow one to algorithmically find relations between
them~\cite{Goncharov:2010jf,Duhr:2011zq,Duhr:2012fh}. One can then
construct a basis for the space of MPLs relevant for five-parton
scattering amplitudes, and this was achieved in~\cite{Gehrmann:2018yef}
where the so-called pentagon functions were introduced.
Further, MPLs are equipped with a notion of weight, which can
be used to organize the pentagon functions. Indeed, there are
no relations between pentagon functions of different weight so we can
separate the space of pentagon functions into subspaces of different 
weights. For two-loop amplitudes, we need functions of at most weight 4.
In the following, we denote the pentagon functions by $\{h_i\}_{i\in B}$ 
and the associated set of labels $B$. 

After expansion in epsilon, the amplitude can thus be expressed in terms
of pentagon functions
\begin{equation}
  \mathcal{A}^{(2)}=
    \sum_{i\in B}\sum_{k=-4}^0\epsilon^{k}
    d_{k,i}h_i\,,
\end{equation}
where we use the fact that the poles in $\epsilon$
of two-loop amplitudes are at most $\mathcal{O}(\epsilon^{-4})$ and
the $d_{k,i}$ are rational functions of the external data.
 The motivation for this decomposition is
that order by order in epsilon the master integrals satisfy more
relations than integration-by-parts relations and these are manifested by the pentagon
function decomposition.
The same basis $B$ can be used to express the terms
${\bf I}_{[n]}^{(1)}\mathcal{A}_R^{(1)}$
and ${\bf I}_{[n]}^{(2)}\mathcal{A}_R^{(0)}$ in 
eq.~\eqref{eq:remaindeDef}, and we can thus decompose the remainder
in terms of pentagon functions:
\begin{equation}
\mathcal{R}^{(2)} = \sum_{i \in B} r_i h_i\,.
\label{eq:remainderPentagon}
\end{equation}
Again, we suppress the dependence of the 
algebraic coefficient functions $r_i$ and the pentagon functions $h_i$ on
the external data.
The remainders, and consequently the coefficient 
functions $r_i$, depend on particle types and helicities as well as
the number of flavors $\NF$.

For convenience we use the convention that the set of pentagon functions $\{h_i\}_{i\in B}$
not only includes genuine functions, such as $\log(-s_{12})$, but also
constants with weight, such as $\pi^2$, that correspond to the pentagon 
functions evaluated at specific points. These are boundary conditions that
are specific to the kinematic region where the amplitude is evaluated. In
this paper we specialize our discussion to the Euclidean region.
In  ref.~\cite{Gehrmann:2018yef}, the pentagon functions have been continued
to all relevant physical regions, and it is thus rather  straightforward
to extend our results to any physical region.

\subsection{Analytic Properties of Coefficient Functions}\label{sec:coefficients}

The coefficient functions $\{r_i\}_{i\in B}$ introduced in 
eq.~\eqref{eq:remainderPentagon}  have a universal pole 
structure~\cite{Abreu:2018zmy} which we find to be expressible in terms of
the so-called planar alphabet $A$ of the pentagon functions \cite{Gehrmann:2018yef}.
The planar alphabet $A$ is given in terms
of 26 letters $\{W_i\}_{i\in A}$ which in turn are rational functions of the independent 
Mandelstam variables $s_{ij}=(p_{i}+p_j)^2$, 
\begin{equation}\label{eq:mandelstams}
\vec s=\{s_{12},s_{23},s_{34},s_{45},s_{51}\}\,,
\end{equation}
and the parity-odd contraction of four 
momenta 
\begin{equation}\label{eq:tr5Def}
\trFive= 4 i \varepsilon^{\mu\nu\rho\sigma}
p_{1\mu} p_{2\nu} p_{3\rho} p_{4\sigma}\, ,
\end{equation}
with the Levi-Civita symbol $\varepsilon_{\mu\nu\rho\sigma}$ and the convention $\varepsilon_{0123}=1$. 
The square of $\trFive$ gives the 
five-point Gram determinant $\Delta_5=\trFive^2$.
The letters can be further grouped into parity even and odd letters, 
$A^+$ and $A^-$ respectively, with $A=A^+\cup A^-$. (Parity here refers to  
the transformation properties of the letters under a parity transformation in 
momentum space.)
While $\trFive$ is parity odd, all Mandelstam variables $\vec s$ are parity even. 
The functions $r_i$ are rational functions in the variables ${\vec s}$ and $\trFive$,  
whose denominators are monomials in the letters,
\begin{equation}
    r_i = \frac{n_{i}}{ W^{\vec q_i}}\,.
    \label{eq:letterDecomposition}
\end{equation}
The vector of exponents $\vec q_i=\{q_{i,1}, ... q_{i,26}\}$ 
differs between the coefficient functions, and for a given
coefficient function $r_i$ not all letters contribute. The pattern of
contributing letters is linked to the pentagon functions and is helicity
dependent~\cite{Abreu:2018zmy}. 
We will often refer to the set of 
independent factors in a monomial such as $W^{\vec\alpha}$ as $A_\alpha$,
\begin{equation}\label{eq:A_alp}
		A_\alpha = \{ i\,|\, i \in A \quad \mbox{with} \quad \alpha_i\neq 0 \} \,.
\end{equation}

Finally, we will have to specify the notion of a monomial ordering 
on the exponent vectors,
\begin{equation}\label{eq:orderingNotation}
 	W^{\vec\alpha} < W^{\vec\beta}  \quad 
 	\leftrightarrow \quad \vec\alpha < \vec\beta \,.
\end{equation}
We will often use the lexicographic monomial ordering, which amounts to comparing 
the size of the leading entries in the vectors and, if they are equal, comparing
the following entries etc.
We refer to text books such as ref.~\cite{cox2013ideals} for further information 
concerning the concepts of monomial ordering in the context of polynomial-division 
algorithms.  

\subsection{Amplitude Evaluation}\label{sec:implementation}

First we summarize the tools employed to obtain the results we present.
The propagator structures $\Delta$ in eq.~\eqref{eq:AL} are obtained
by generating colored cut diagrams with \texttt{QGRAF} \cite{Nogueira:1991ex}
followed by color decomposition performed in \Mathematica{}
 according to
\cite{Ochirov:2016ewn,FermionColour}.
We carry out the decomposition into master and surface integrands
as described in refs.~\cite{Abreu:2017xsl, Abreu:2017hqn, Abreu:2018jgq,
Abreu:2018zmy}, where we used SINGULAR
\cite{DGPS} to obtain the unitarity-compatible surface terms.
The master integral coefficients in eq.~\eqref{eq:AL} are evaluated over finite 
fields,\footnote{With cardinalities of order $\mathcal{O}(2^{31})$.}
employing our \texttt{C++} framework for multi-loop numerical unitarity.
We use \texttt{Givaro} \cite{Givaro} for basic finite-fields arithmetic, and 
improve the multiplication speed with a custom implementation
of a Barrett reduction \cite{Barrett1987,HoevenLQ14}. The $D_s$-dimensional
tree-level amplitude products appearing on the left-hand-side of
eq.~\eqref{eq:CE} are evaluated
through off-shell recursion~\cite{Berends:1987me}, and the corresponding linear
system of equations are solved by PLU factorization and back substitution.

Furthermore, we document some technical aspects of the computations.
The transformation from master integral coefficients in eq.~\eqref{eq:Amaster}
into the coefficients of pentagon functions in eq.~\eqref{eq:remainderPentagon}
is accomplished with our \texttt{C++} representation of the master integrals in
terms of pentagon functions~\cite{Gehrmann:2018yef}.
We refine the treatment of square roots appearing from solving the quadratic 
on-shell conditions on the loop momenta
compared to the previous implementation described in ref.~\cite{Abreu:2018jgq}.
As before, we rotate the $D$-dimensional components of the loop momenta 
into a six-dimensional subspace,
\begin{equation}
  \ell_1 = (\ell_{1[4]},\vec{\mu}_1)\,, \qquad \ell_2 = (\ell_{2[4]},\vec{\mu}_2)\,.
  \label{eq:loopmomenta}
\end{equation}
We first use the algorithm of~\cite{Abreu:2017hqn} to solve the on-shell
conditions such that the $\mu_{ij}$ are rational in the input
parameters. In order to represent these momenta in a 6-dimensional
embedding we proceed as follows.  Without loss of generality, we choose
to use an alternating metric signature $(+,-,+,-,\ldots)$, and
parametrize the two-dimensional loop-momentum components 
$\vec{\mu}_1$ and $\vec{\mu}_2$ as follows:
\begin{equation}
  \vec{\mu}_1(t)  = \frac{1}{2}\begin{pmatrix}
    t+\dfrac{\mu_{11}}{t} \\
    t-\dfrac{\mu_{11}}{t} \\
  \end{pmatrix}, \quad
  \vec{\mu}_2(t)  = \frac{\mu_{12}}{\mu_{11}}\vec{\mu}_1(t) - \frac{r}{\mu_{11}}~\frac{1}{2}\begin{pmatrix}
    t-\dfrac{\mu_{11}}{t} \\
    t+\dfrac{\mu_{11}}{t} \\
  \end{pmatrix},
  \label{eq:muparam}
\end{equation}
where $t$ is a free dimensionful parameter that leaves the scalar products
$r = \sqrt{\mu_{12}^2-\mu_{11} \mu_{22}}$ and 
$\mu_{ij} = \mu_i^1 \mu_j^1 - \mu_i^2 \mu_j^2$ invariant. %
We perform numerical computations with the external kinematic data 
$\{ p_i\}_{i=1,5}$ taking values in a finite field. While we do not
require loop momenta to take values in the same number field,
eq.~\eqref{eq:muparam} guarantees that their components take values in an
algebra generated by the basis $\{1,r\}$ over the same number field.
The above parametrization is an improvement compared to the 
basis of four elements employed in ref.~\cite{Abreu:2018jgq}.

\section{$D_s$ Dependence from Dimensional Reduction}\label{sec:Ds}

The amplitudes $A$ defined in eq.~\eqref{eq:defAmpsTens} are polynomials in $D_s$,
\begin{equation}
  A(D_s) = \sum_{i=0}^{N} \mathcal{K}_i~D_s^{i}\ ,
  \label{eq:ds-poly}
\end{equation}
where $N$, the maximal power of $D_s$, 
varies depending on the process, the loop order, and the choice of tensor structure 
in eq.~\eqref{eq:tensorDecomposition}.
For the amplitudes considered in this paper $N\leq2$.
In this section we will suppress all arguments of $A$ and only keep track of the 
dependence on~$D_s$.
In a numerical framework, $A(D_s)$ can only be evaluated for integer $D_s$ values for which 
the particle states are well defined.
To be able to set $D_s=4-2\epsilon$ in the HV scheme, the knowledge of the coefficients $\mathcal{K}_i$ is required.
One way to obtain them is to reconstruct the polynomial \eqref{eq:ds-poly} from
a sample of $(N+1)$ integer values of $D_s$.
This procedure is known as dimensional reconstruction \cite{Giele:2008ve} and has  previously been applied
in \cite{Ellis:2008ir,Boughezal:2011br,Abreu:2017xsl,Abreu:2017hqn}.

While being generic and straightforward to implement, this approach has drawbacks which
become particularly evident in amplitudes with fermions.
The dimension of the spinor representation scales exponentially (as $2^{D_s/2}$) with $D_s$,
as opposed to the linear scaling of the vector representation.
Furthermore, the external spinor states with definite helicity can be embedded consistently
only for even values of $D_s$,
which pushes the sample values higher compared to the case of vector particles in the loops.
Beyond the obvious detrimental effect on the numerical complexity, the dimensionality of the
spinor representation determines the number of terms entering the evaluation of the traces
to obtain the helicity amplitudes through  eq.~\eqref{eq:defAmpsTens}. For
the case of amplitudes with multiple external quark pairs this makes the computation
of traces unnecessarily time consuming.

These considerations motivate the search for more efficient alternatives to dimensional
reconstruction. Here we employ one such alternative, based on the idea of
dimensional reduction, which has recently been presented in ref.~\cite{Anger:2018ove} 
and already applied to the computation of one-loop amplitudes in ref.~\cite{Anger:2017glm}.
In the remainder of this section we give a brief overview of this method 
and refer the reader to ref.~\cite{Anger:2018ove} for more technical details.

We start by rearranging eq.~\eqref{eq:ds-poly} in the following way:
\begin{equation}
  A(D_s) = \sum_{i=0}^{N} \tilde{\mathcal{K}}_i~(D_s-D_0)^{i},
  \label{eq:ds-poly-alt}
\end{equation}
where $D_0$ is some \textit{base} dimension, and the coefficients $\tilde{\mathcal{K}}_i$
can be obtained by a linear transformation of the coefficients $\mathcal{K}_i$ in eq.~\eqref{eq:ds-poly}.
It turns out that, given a suitable choice of $D_0$,
the dependence of $A(D_s)$ on degrees of freedom higher than $D_0$ can
be captured in a kinematic-independent way.  
This observation allows one to analytically separate this
dependence, and thus evaluate each coefficient $\tilde{\mathcal{K}}_i$
directly. Furthermore, these evaluations are then performed in the base
dimension $D_0$, resulting in spinor representations of much lower dimensionality
compared to those encountered in the framework of dimensional reconstruction.

For reasons that will become clear shortly, one chooses the base
dimension $D_0$ to be the minimal dimension which allows to embed 
all loop-momentum components without introducing new relations.
For two-loop amplitudes we have $D_0=6$, and we
shall specialize to this case henceforth. We write the metric tensor as
a direct sum,
  \begin{align}
    \label{eq:ds-split-metric}
    g^{\mu\nu}_{[D_s]}  = g^{\mu\nu}_{[D_s-6]} + g^{\mu\nu}_{[6]},  \qquad
    g^{\mu\nu}_{[D_s-6]}g^{\phantom{\mu\nu}}_{\mu\nu\,[6]} = 0\,,
  \end{align}
  and the gamma matrices as a direct product 
  (see e.g.~\cite{Collins:1984xc,Kreuzer:susylectures}),
\begin{equation}
  (\gamma_{[D_s]}^\mu)_{a\kappa}^{\,b\lambda}  = \left\{ 
    \begin{array}{ll} 
      \left(\gamma_{[6]}^\mu\right)_a^{\;b} \,
      \delta_\kappa^\lambda\,, &\quad  0\le\mu \le 5 \,,\\&\\
      \left(\gamma^\star_{[6]}\right)_a^{\;b} 
      \left(\gamma_{[D_s-6]}^{(\mu-6)}\right)_\kappa^{\;\lambda}\,, 
      &\quad \mu \geq 6 \,,
    \end{array}
    \right.
    \label{eq:ds-gamma}
\end{equation}
where $\gamma^\star_{[6]}$ is a six-dimensional analogue of $\gamma_5$ in four dimensions, i.e.
$\{\gamma^\star_{[6]},\gamma_{[6]}^\mu\} = 0$ for all $\mu \in \{0,5\}$, and $(\gamma^\star_{[6]})^2 = 1$.
The product representation allows us to factorize any chain of gamma matrices into a product of $6$- and $D_s-6$-dimensional gamma-matrix chains.
Then, using the fact that the trace of a direct product of two matrices is the product of their traces,
we can split the traces required to obtain the coefficients of tensor structures in eq.~\eqref{eq:tensorDecomposition}
as follows:
\begin{equation}
  \Tr\left(\prod_{\mu_i\in \mathcal{G}}\gamma^{\mu_i}_{[D_s]}\right) =
  \Tr\left(\prod_{\mu_i\in \tilde{\mathcal{G}}}\gamma^{\mu_i}_{[D_s-6]}\right) \cdot
  \Tr\left(\prod_{\mu_i\in \mathcal{G}}\gamma^{\mathfrak{I}(\mu_i)}_{[6]}\right),
  \label{eq:ds-split-traces}
\end{equation}
where the product on the left-hand side is over a sequence $\mathcal{G}$ of $D_s$-dimensional Lorentz indices,
$\tilde{\mathcal{G}} = \{ \mu_i \in \mathcal{G} ~\vert~ \mu_i \geq 6 \}$, and the map $\mathfrak{I}$ is defined as
\begin{equation}
  \mathfrak{I} : \mu \to
    \begin{cases}
      \mu, & 0\le\mu \le 5\,, \\
      \star, & \mu \geq 6\,.
    \end{cases}
\end{equation}
The traces of $\prod_{\mu_i\in\tilde{\mathcal{G}}}\gamma^{\mu_i}_{[D_s-6]}$ can be 
evaluated analytically using well-known Clifford algebra identities,
which produce sums of products of $g^{\mu_i\mu_j}_{[D_s-6]}$.
The crucial observation is that the only object to be contracted with the indices beyond 
$(D_s-6)$ is $g^{\mu\nu}_{[D_s-6]}$.
This is ensured by our choice of the base dimension.
These indices then always contribute terms of the form $g^\mu_{[D_s-6]\mu} = (D_s-6)$, 
generating contributions to the coefficients $\tilde{\mathcal{K}}_i$ with $i>0$ 
in eq.~\eqref{eq:ds-poly-alt}. At this point, all degrees of freedom beyond $(D_s-6)$
are traded for polynomials in $(D_s-6)$ with integer factors, and the coefficients
$\tilde{\mathcal{K}}_i$ are expressed in terms of six-dimensional objects only.

From a Feynman diagrammatic perspective, the contributions to the
coefficients of the polynomial in $(D_s-6)$
can be represented by introducing a scalar particle. 
The Feynman rules associated to this particle can be readily
derived from dimensional reduction of the original Feynman rules of the
theory \cite{Bern:2002zk,Badger:2013gxa,Giele:2008ve,Anger:2018ove} by
applying the relations \eqref{eq:ds-split-metric} and \eqref{eq:ds-gamma}.
We list them in the appendix \ref{sec:DsFeynRules} for convenience.
To illustrate this procedure we now give an example. 
Consider a decomposition of a Feynman diagram with $D_s$-dimensional particles on the 
left-hand side of figure~\ref{fig:ds-example-diagram}. The four non-vanishing 
contributions after evaluating partial traces and contracting all 
$(D_s-6)$-dimensional indices are shown on the right-hand side, 
where the scalars are introduced to represent what remains of these contractions.

\begin{figure}
  \newlength\diagramWidth
  \setlength{\diagramWidth}{18ex}
  \centering
  \begin{align*}
    \vcenter{\hbox{\includegraphics[width=\diagramWidth]{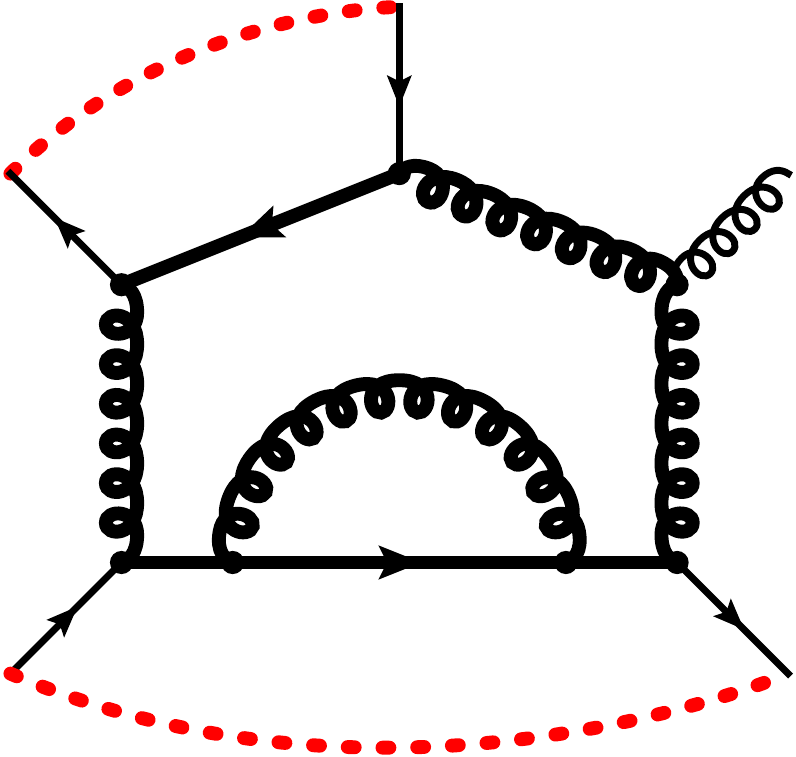}}} \hspace{-10ex} & \hspace{15ex} = \hspace{6ex}
    \vcenter{\hbox{\includegraphics[width=\diagramWidth]{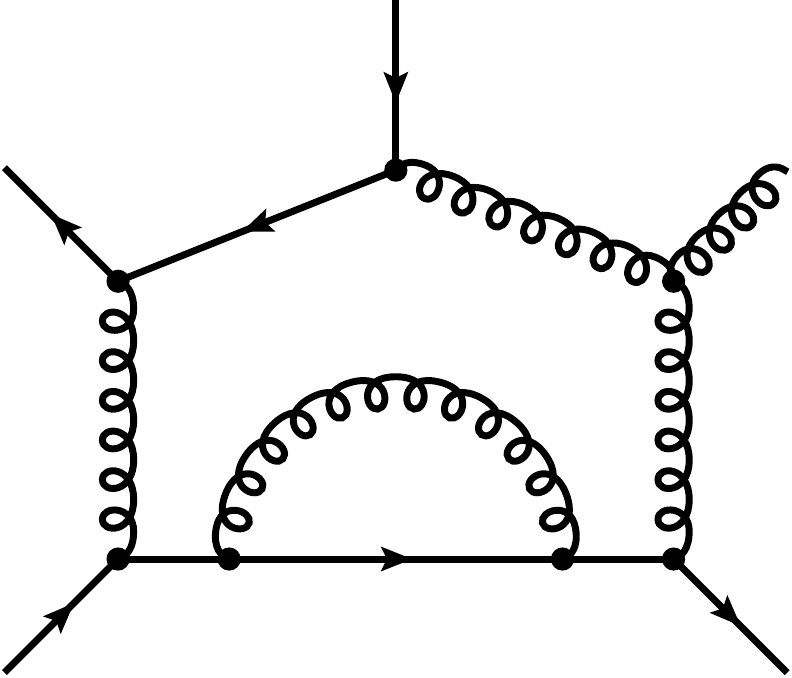}}} \quad +\\
    \Big(~ &
    \vcenter{\hbox{\includegraphics[width=\diagramWidth]{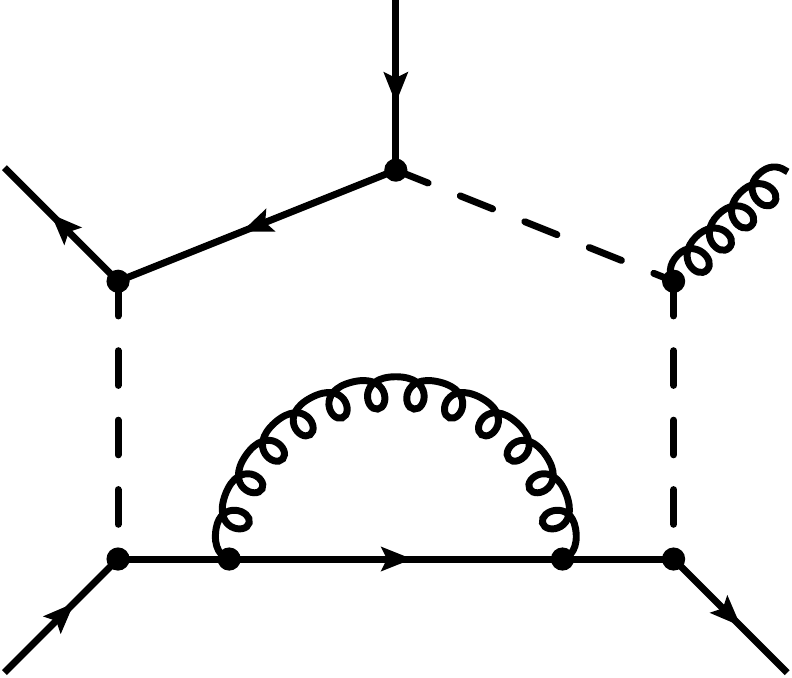}}} \quad+\quad
    \vcenter{\hbox{\includegraphics[width=\diagramWidth]{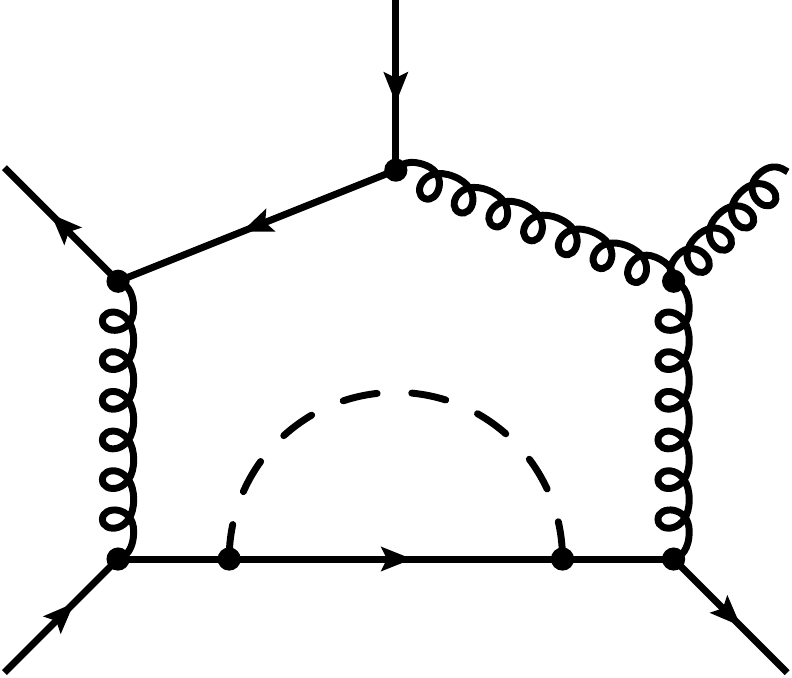}}}
    ~\Big) ~\cdot~ (D_s-6) \quad +\\
    &
    \vcenter{\hbox{\includegraphics[width=\diagramWidth]{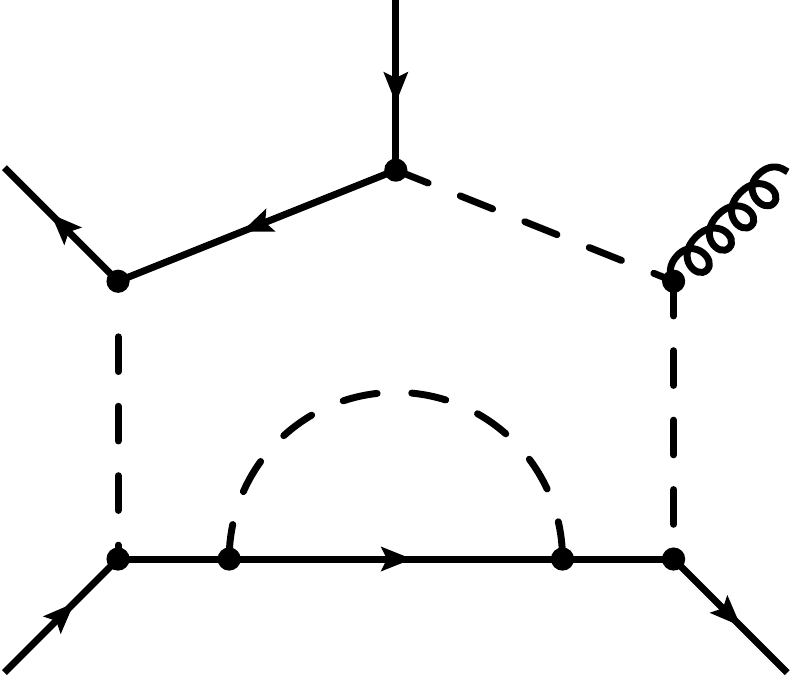}}}
    ~\cdot~ (D_s-6)^2,
  \end{align*}
  \caption{Example of diagrams with scalar particles,
    representing the contributions to the coefficients of $\tilde{\mathcal{K}}_i$ in eq.~\eqref{eq:ds-poly-alt}.
    The thick lines in the diagram on the left-hand side represent particles in arbitrary $D_s$ dimensions.
    The (red) dashed lines connecting external-quark lines represent the traces required to obtain the coefficient of the tensor structure of eq.~\eqref{eq:defAmpsTensb}.
    All particles on the right-hand side are in six dimensions.}
  \label{fig:ds-example-diagram}
\end{figure}

We would like to conclude this section with some remarks.
First, the remaining non-trivial traces of eq.~\eqref{eq:ds-split-traces},
i.e.~those of $\prod_k\gamma^{\mu_k}_{[6]}$, cannot be simplified generically
as the indices appear contracted with the loop momenta at the integrand level.
We evaluate them by the direct summation over a specially constructed set of external states
(in an analogous way to the sums performed in~\cite{Abreu:2018jgq} 
for dimensional reconstruction).
Secondly, in the absence of fermions in the loops, this method coincides with the so called
six-dimensional formalism employed in refs.~\cite{Badger:2013gxa,Badger:2017jhb}, and can 
thus be viewed as an extension thereof to amplitudes with fermions.
Finally, we note that the method presented in this section can be straightforwardly
generalized to higher number of loops by adjusting the base dimension $D_0$,
as well as to the extraction of coefficients of different tensor structures in 
eq.~\eqref{eq:tensorDecomposition}.

\section{The Analytic Structure of Five-Parton Amplitudes}\label{sec:reconstruction}

We apply three methods to reduce the complexity of the remainder
functions and facilitate their reconstruction from numerical samples.
First, we introduce a particular ansatz for the coefficients $r_i$ of 
eq.~\eqref{eq:remainderPentagon} which simplifies the reconstruction procedure by
decreasing the required number of sample points. 
Next, we comment on how to remove redundancies
in the large number of rational functions in a scattering amplitude. 
Finally, we reduce the difficulty of rationally reconstructing the finite-field 
result by introducing an algorithm to uniquely obtain a
multivariate partial-fraction decomposition of the analytic expressions.

\subsection{Analytic Reconstruction}
\label{sec:reconstructionInvs}

In order to perform the reconstruction of the five-parton amplitudes from
samples over finite fields, we employ a modification of the algorithm of
\cite{Peraro:2016wsq}. 
Our aim is to reconstruct the functions $r_i$ in terms of $\{ \vec s, \trFive \}$,
with $\vec s$ the five independent Mandelstam invariants defined in
\eqref{eq:mandelstams} and the $\trFive$ defined in \eqref{eq:tr5Def}, 
while using a suitable rational parametrization of the 
phase space of five massless particles.

Let us start by introducing  such a parametrization of 
the external kinematics. We keep four of the Mandelstam invariants
$\{ s_{12}, s_{23}, s_{45}, s_{51} \}$ and introduce an auxiliary variable $x$.
More explicitly, we consider the Mandelstam invariants as functions of
this new set of variables,
\begin{equation}
  \vec s\equiv \vec s\,(s_{12}, s_{23}, s_{45}, s_{51},x)\,,
\end{equation}
and, rescaling the momenta such that $s_{12}=1$, $s_{34}$ is defined in terms of
$x$ as
\begin{align}
\begin{split}
  s_{34}\equiv
    s_{34}(s_{23}, s_{45} , s_{51}, x) =& \frac{ (s_{45} - 1) s_{51} x -
	s_{23} (s_{23} -  s_{45} - s_{51} - x)}{(s_{45} - s_{23} + x)\,x}\,.
    \label{eq:s34Definition}
\end{split}
\end{align}
This parametrization follows
from the twistor matrix given in appendix~\ref{sec:twistorVariables} and so naturally
rationalizes $\trFive$.
For any (generic) fixed value of the invariants $\vec s$
there are two corresponding values of $\{x,\bar x\}$, corresponding to the 
two solutions to the quadratic equation in \eqref{eq:s34Definition}.
This manifests the known fact that it is not possible to rationally parametrize the phase space in
terms of Mandelstam variables.
These two different points in phase space are in fact parity conjugates, 
and correspond to opposite signs of $\trFive$ (which is the square root of the 
discriminant of the quadratic equation in \eqref{eq:s34Definition}). 
Under this transformation,
\begin{align} 
  \label{eq:parity}
  \trFive\to-\trFive\,,\qquad
  x\to\bar x=
  \frac{s_{23}(s_{23}-s_{45}-s_{51})}{s_{34}(s_{12}, s_{23}, s_{45}, s_{51},x)\,x}\,.
\end{align}

Having established the set of variables that rationalize the 
phase space of five massless particles, we next
discuss our ansatz for the rational coefficients $r_i$
and the sampling procedure to fix the parameters in the ansatz. 
We  decompose each function $r_i$
into parity odd and even parts,
\begin{equation}
    \label{eq:ripmdef}
		 r_i(\vec s, \trFive ) = r_i^+(\vec s\,)  + \trFive \, r_i^-(\vec s\,)\,. 
\end{equation}
As $\trFive{}^2$ is manifestly polynomial in the invariants, it is clear that
no higher powers in $\trFive$ are required. 
The motivation for choosing such a
representation is two-fold. First, a generic expression of this form
has a higher polynomial degree when expressed in terms of the twistor
variables, as the relation is non-linear. 
Consequently, as we shall observe, expressing
the finite remainders in this form
allows for a more efficient analytic reconstruction.
Second, the Mandelstam variables 
make manifest the dependence on
the external momenta, and therefore also any possible symmetries related to 
exchanges of external legs.  

While the functions $r_i^{\pm}$  are rational in the Mandelstam 
invariants $\vec s$, their
definition through eq.~\eqref{eq:ripmdef} only allows them to be
evaluated via the functions $r_i$ which, due to the presence of
$\trFive$, are not rational in the invariants themselves. 
For this reason, we first isolate the even and odd parts $r_i^\pm$
and reconstruct them in terms of the Mandelstam variables $\vec s$.
As the parity conjugate points $\{s_{23},s_{45},s_{51}, x\}$
and $\{s_{23},s_{45},s_{51}, \bar x\}$ 
correspond to  the same  value of the invariants $\vec s$, we can use
this pair of points to evaluate the $r_i^{\pm}$ on a single point~$\vec s$,
\begin{eqnarray}
		r^+_i(\vec s\, ) &=& \frac{1}{2} \big[ r_i(\vec s, x ) + r_i(\vec s,\bar x ) \big]\,,\\
		r^-_i(\vec s\, ) &=& \frac{1}{2\, \trFive } \big[ r_i(\vec s,x ) - r_i(\vec s,\bar x ) \big]\,.
\end{eqnarray}

A further simplification to the reconstruction procedure 
is achieved by conjecturing that the ansatz for the
denominators given in eq.~\eqref{eq:letterDecomposition} extends to the $r_i^{\pm}$,
with the extra requirement that no odd letters appear in the denominator.
That is, we conjecture that
\begin{equation}
    r^\pm_i(\vec s\,)  = \frac{n_i^\pm(\vec s\,) }{W^{\vec q_i}(\vec s\,) }\,,
   \label{eq:DenominatorAnsatz}
\end{equation}
where the $n_i^\pm(\vec s\,)$ are polynomials in the invariants and
the contributing $W_j(\vec s\,)$ are polynomials in the $\vec s$ only (and not $\trFive$). 
To test this
ansatz and determine the exponent vector $\vec q_i$, we consider the functions
$r_i^{\pm}$ on a `univariate slice' where the variables depend on a single parameter 
$t$, $\{s_{23}(t),s_{45}(t),s_{51}(t), x(t)\}$. 
We then reconstruct the univariate functions $r_i^{\pm}(t)$ using Thiele's 
method and match the denominators 
with a monomial in the letters $W_i(\vec s\,[t])$ in order to obtain the exponent vector $\vec q_i$.
In order to recover this information from the univariate
slice, we must choose a curve on which all 
letters $W_i(\vec s\,[t])$ are distinct functions of $t$.
Furthermore, if we require that all invariants are linear in $t$, this 
procedure will also tell us the total degree of the numerator polynomials  
$n_i^\pm(\vec s\,)$ in the Mandelstam variables. Such a curve is given by, e.g., 
\begin{equation}
	x(t) = c_0, \quad  s_{23}(t) = c_1 + d_1 t, \quad s_{45}(t) = c_2 + d_2 t, \quad  s_{51}(t) = c_3 \big(x(t) - s_{23}(t) + s_{45}(t) \big)\,,
\end{equation}
where the
variables $c_i$ and $d_i$ take fixed, arbitrary values.\footnote{It is much more transparent in the
parametrization of \cite{Abreu:2018zmy} as to why $s_{34}(t)$ is then linear in $t$.}
With this procedure, we confirmed the ansatz of eq.~\eqref{eq:DenominatorAnsatz}
and thus determined all the denominators of the rational functions $r_i^{\pm}$.

The problem of reconstructing the $r_i^{\pm}$ is now reduced to the
reconstruction of the polynomials $n^\pm_i$.  We perform the analytic reconstruction
of $n_i^\pm(\vec s)$ directly in terms of the Mandelstam variables. 
This is achieved with a modified form
of the recursive Newton method \cite{Peraro:2016wsq}. If we consider a
univariate polynomial in the variable $z$, the method makes an ansatz which is
adapted to the choice of sample points $\{z_1, \ldots, z_n\}$. Specifically,
the polynomial is written as a linear combination of the basis polynomials
$$p_j(z) = \prod_{i=0}^{j-1} (z - z_i).$$ 
As the $p_j(z_k)$ vanish for $k<j$, when solving for the coefficients of this
basis the linear system is triangular by construction. Once one establishes
the coefficients of these basis elements, it is trivial to rewrite the polynomial
in terms of monomials in $z$.  In the multivariate case, one can apply this
strategy recursively by singling out one variable and letting the coefficients
be polynomials in the remaining variables \cite{Peraro:2016wsq}.

In our approach we also use the observation that it is not strictly required that 
the arguments of the $p_j$ be the input parameters. 
In fact, we can choose them to be a function, in our case 
$s_{34}=s_{34}(s_{23}, s_{45}, s_{51}, x)$.
Specifically, in the first step of the recursion, we write our numerator
polynomial as
\begin{equation}
    n_i^{\pm}(s_{23}, s_{45}, s_{51}, s_{34}(s_{23}, s_{45}, s_{51}, x)) = 
	\sum_{j = 0}^R n_{i,j}^{\pm}(s_{23}, s_{45}, s_{51}) 
  p_j(s_{34}(s_{23}, s_{45}, s_{51}, x)),
   \label{eq:polynomialMandelstamAnsatz}
\end{equation}
where
\begin{equation}
p_j(s_{34}(s_{23}, s_{45}, s_{51}, x))
=\prod_{i=0}^{j-1}(s_{34}(s_{23}, s_{45}, s_{51}, x)-
s_{34}(s_{23}, s_{45}, s_{51}, x_i))\,,
\end{equation}
the value $R$ is obtained from the highest degree term $\sim t^R$ found
in the univariate slice, and the $n_{i,j}^{\pm}$ are polynomials. 
To be able to evaluate the $n_{i,j}^{\pm}$ at a point in the recursive approach, 
we evaluate the function $n_i^\pm$ at $R+1$ values of $x$, and solve the associated
triangular linear system.

To illustrate the advantage of the approach outlined above compared with the
strategy used in \cite{Abreu:2018zmy}, we show in
figure~\ref{fig:degrees-twXsij}  the polynomial-degree drop for the
two most complex remainders, separating the different $\NF$ contributions.
This observation is common to all five-parton remainders at $N_f^0$ and $N_f^1$: 
we observe a drop of the polynomial degree in the numerators  $n_i^\pm(\vec s)$ by 
40\%-50\% when written in terms of Mandelstam variables as compared to the twistor 
variables.
The efficiency of the analytic reconstruction algorithm scales as a binomial
$\left(\substack{n+R \\ n }\right)$ in the number of variables $n$ and the
polynomial degree $R$. Such a drop in the polynomial degree thus 
implies an important improvement in efficiency. 
The $N_f^2$ remainders are trivial to reconstruct in both approaches.

Finally, as an indication for the complexity of the analytic reconstruction we show the
numerator degrees for the $N_f^0$ contribution of all the planar five-parton
remainders in figure~\ref{fig:degrees-all}. The figure displays the expected reduction in
complexity of the remainders for external fermions compared to gluons.

\begin{figure}
  \centering
  \includegraphics[width = 0.7\textwidth]{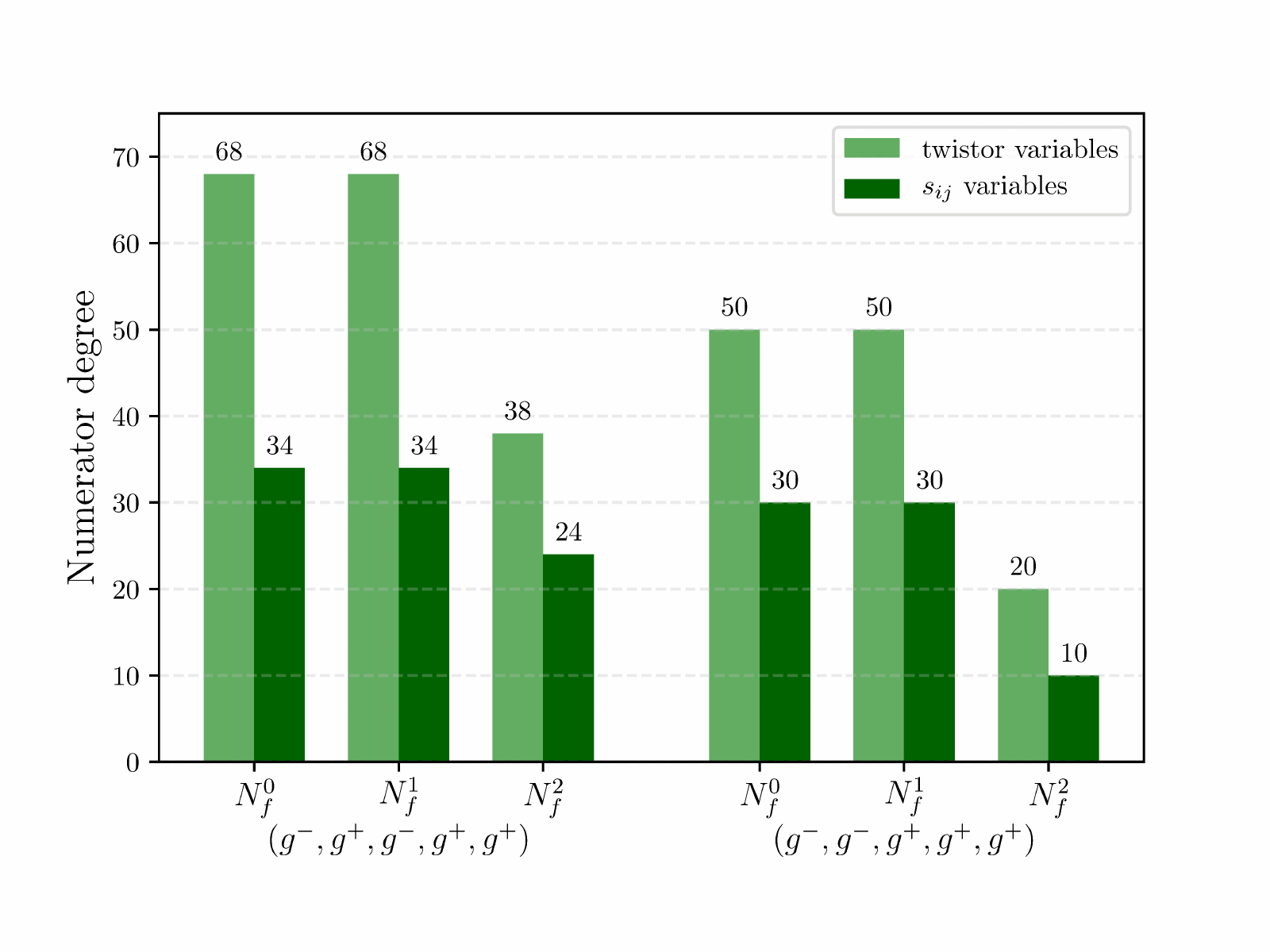}
  \caption{Maximal degrees of numerators of the two most complicated amplitudes in twistor and $s_{ij}$ variables.}
  \label{fig:degrees-twXsij}
\end{figure}

\begin{figure}
  \centering
  \includegraphics[width = 0.7\textwidth]{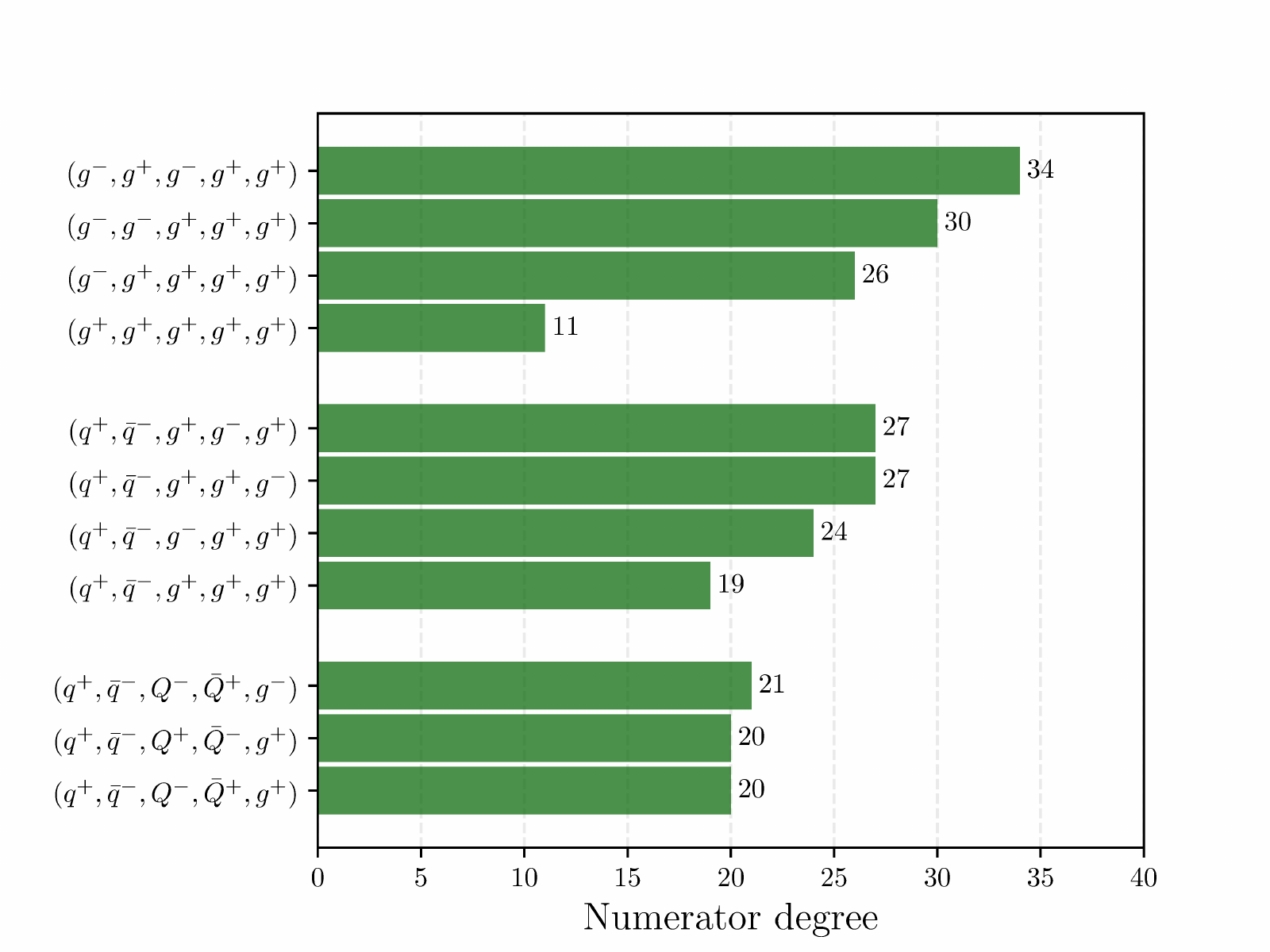}
  \caption{Maximal degrees of numerators of $N_f^0$ components of the amplitudes in $s_{ij}$ variables.}
  \label{fig:degrees-all}
\end{figure}

\subsection{Basis of Rational Coefficients}

It was noted in \cite{Abreu:2018aqd,Chicherin:2018yne,Abreu:2019rpt,Chicherin:2019xeg} 
that the space of independent rational
functions appearing in the $\mathcal{N}=4$ super Yang-Mills (SYM) and $\mathcal{N}=8$ super gravity
two-loop five-point amplitudes is much smaller than the space of possible independent
pentagon functions. This motivates us to apply the same analysis here, both
in order to allow performing minimal reconstruction work and as a way to obtain
simpler analytic expressions.
These linear dependencies between the rational functions allow us to express the 
remainder of each amplitude as
\begin{equation}
		\mathcal{R}^{(2)} = 
		\sum_{i\in K ,\, j\in \hat B} r_i M_{ij} h_j,
\label{eq:remainderDecomposition}
\end{equation}
where $r_i$ is a vector of rational coefficients and $K$ denotes
the dimension of the space they span.
The index $j$ runs over a subspace $\hat B$ of the space $B$ of all pentagon functions 
introduced in eq.~\eqref{eq:remainderPentagon}, i.e., $\hat B\subseteq B$. 
The constant matrix $M$, the rational functions $r_i$ 
and the index sets $K$ and $\hat B$ are helicity-amplitude dependent.

The basis of rational functions $r_i$ is not unique
and we choose it by taking the linearly independent subset with
the lowest total polynomial degree. The associated matrix $M$
can be computed numerically in the finite field used for the analytic reconstruction,
using as many evaluations as its rank. 
In order to rationally reconstruct $M$ we only
require one finite field of cardinality $\mathcal{O}(2^{31})$, apart from two
cases where we combine information from two finite fields (see more details in 
section~\ref{sec:results}).

\subsection{Partial Fractions}
\label{sec:PartialFractions}

Having chosen a basis of rational coefficients, we study the $\{r_i^\pm\}_{i\in K}$
in a finite field, obtained with the strategy described 
in section \eqref{sec:reconstructionInvs}.
The next step is to lift the finite-field result to the rational numbers, 
i.e., to rationally reconstruct the result. 
The correctness of the rational reconstruction can
be verified by comparison against a single evaluation on a
different finite field. 
In general, the success of the rational reconstruction is
guaranteed if the numerator and denominator of the original rational
number satisfy a bound dependent on the finite-field cardinality
\cite{Wang:1981:PAU:800206.806398, vonManteuffel:2014ixa, Peraro:2016wsq},
i.e., rational numbers with numerators and denominators with
smaller absolute values are more easily reconstructed from their finite-field
images. As such, it is beneficial if we can find a form of the
$r_i^\pm$ where the rational numbers involved have `simple' numerators and
denominators. 
To this end, we will employ a multivariate partial-fraction decomposition
of the $r_i^\pm$, which will also have the side effect of simplifying their
analytic form as a whole. 
Indeed, the partial-fraction decomposition manifests the singularities
of the coefficients, which are controlled by physical properties such
as the factorization of the amplitude in specific limits.

We will employ the partial-fraction decomposition introduced by
Le\u{\i}nartas \cite{leinartas1978factorization,raichev2012leinartas}.
This decomposition has recently been used for bringing differential equations into
canonical form~\cite{Meyer:2016slj}, as well as for the
analytic reduction of loop integrands~\cite{Zhang:2012ce, Mastrolia:2012an}.
A Le\u{\i}nartas decomposition of our basis functions is given by
\begin{equation}
    r_i^{\pm} = \sum_{\vec\alpha \in D_i} \frac{n^{\pm}_{i,\vec\alpha}}{ 
	W^{\vec\alpha} }\,.
    \label{eq:LeinartasDecomposition}
\end{equation}
That is, the $r_i^{\pm}$ are written as a linear combination of rational
functions which we label by the exponents $\alpha$ of the denominators.
The set of all $\vec\alpha$ for a given $r_i^{\pm}$ is denoted by $D_i$ (we combine
the exponents for the even and odd coefficients). The 
$\{n_{i,\vec\alpha}^\pm \}_{\vec\alpha\in D_i}$ are the numerators associated
with the rational function labelled by $\vec\alpha$,
which are polynomials in the $\{\vec s\}$ variables.
We construct the Le\u{\i}nartas decomposition 
\eqref{eq:LeinartasDecomposition} coefficient-by-coefficient, such that the exponents in 
$\vec\alpha\in D_i$ are non-vanishing only for the letters appearing 
in the starting denominator $W^{\vec q_i}$, see eq.~\eqref{eq:DenominatorAnsatz}, 
which we recall are polynomials in the $\vec s$.
The particular algorithm which we employ to uniquely specify the 
Le\u{\i}nartas decomposition is related to the techniques 
of \cite{Smirnov:2005ky}.

A Le\u{\i}nartas decomposition \eqref{eq:LeinartasDecomposition}
is characterized by exponents of the denominator factors. 
All denominators $W^{\vec \alpha}$
in a Le\u{\i}nartas decomposition  are `algebraically independent', in that
there exists no non-zero polynomial  
$P(w)$, such that it vanishes upon inserting the denominator factors, 
\begin{equation}\label{eq:algDep}
P\big(\{ W_i \}_{i\in A_\alpha}\big )=0\,,
\end{equation}
where $A_\alpha$ denotes the set of letters in $W^{\vec \alpha}$, see eq.~\eqref{eq:A_alp}.
Let us make a few comments about such a decomposition.  First, individual terms
in eq.~\eqref{eq:LeinartasDecomposition} may have factors raised to higher
degree than when expressed over a common denominator as in 
eq.~\eqref{eq:DenominatorAnsatz}. Second, no term can have
more denominator factors than the number of variables $\vec s$ as these factors would
necessarily be algebraically dependent.  Also, we note that the Le\u{\i}nartas
decomposition is different to an iterated univariate partial-fraction decomposition in that
it maintains the original denominator structures of the input 
function, i.e., in our case, of eq.~\eqref{eq:DenominatorAnsatz}. Finally, in
ref.~\cite{raichev2012leinartas} it is noted that such a
decomposition is in general not unique, due to how the algebraic dependence relations
\eqref{eq:algDep} are resolved.
Let us illustrate this point by considering
\begin{equation} 
P(w_3,w_4,w_6)=w_3 + w_4 - w_6\,,
\end{equation} 
which clearly vanishes upon 
using the definitions of the letters, 
$W_3=s_{34}, W_4=s_{45} $ and 
$W_6 =s_{34} + s_{45}$. 
The dependence relation implies a relation between multiple terms with different denominator factors,
\begin{equation}\label{eq:arbitraryLeinartas}
0= \frac{P(W_3,W_4,W_6)}{W_3 W_4 W_6} 
= \frac{1}{W_4 W_6} + \frac{1}{W_3 W_6} - \frac{1}{W_3 W_4}\,, 
\end{equation}
which can be used to remove either of the three denominator factors on the right-hand side.
This arbitrariness in the implementation of the Le\u{\i}nartas decomposition will
in general lead to expressions that are not as compact as possible if inconsistent choices
are made across different terms being decomposed.
In the following we describe
an approach that produces a unique Le\u{\i}nartas decomposition by providing 
a global prescription to expand algebraic dependencies of the denominators of a given
coefficient $r_i^{\pm}$.

Our approach makes use of multivariate polynomial division and
Gr\"obner basis techniques. In summary, we reinterpret a rational
function in a polynomial way in order to employ division by a Gr\"{o}bner basis
to find a canonical form of a
polynomial which is subject to a series of constraints. The set of
constraints among the variables form a generating set of an ideal and
if one divides a polynomial in these variables by a Gr\"{o}bner basis of
the ideal, the remainder is unique and `minimal' with respect to the
relations. In the rest of this section we describe these steps in more
detail. We refer the reader to ref.~\cite{cox2013ideals} for a pedagogical
discussion of all the required concepts in polynomial-division algorithms.

Our starting point is a rational function $r_i^\pm$ 
of the form \eqref{eq:DenominatorAnsatz}, and 
we introduce an auxiliary variable for each
denominator factor $W_j$ for $j\in A_{q_i}$, where we recall $A_{q_i}$ denotes
the set of labels of letters appearing in the denominator of $r_i^\pm$.
More precisely we introduce variables $Q_j$ to which we
associate the constraint polynomial $C_j(Q_j, \vec s\,)$,
\begin{align}
    C_j &= W_j (\vec s) \, Q_j - 1\,, 
	\quad\mbox{for all}\quad  j \in  A_{q_i}\,.
   \label{eq:PartialFractionConstraints}
\end{align}
Setting all $C_j = 0$ (i.e., working in the equivalence class of the ideal 
generated by $\{ C_j \}_{j \in A_{q_i}}$)
imposes the constraint that multiplication by $Q_j$
is equivalent to division by $W_j$. 
As such, subject to this constraint we can
express eq.~\eqref{eq:DenominatorAnsatz} in a polynomial fashion as
\begin{equation}
    r_i^{\pm}(\vec{s}\,) \sim n_i^{\pm}(\vec s)\, Q^{\vec q_{i}}\,,
    \label{eq:CartialFractionsLifted}
\end{equation}
where the notation '$\sim$' emphasises that the relation holds modulo 
the ideal $\{ C_j \}_{j \in A_{q_i}}$.
In order to uniquely implement the constraints $C_j = 0$, we first divide
the right-hand side of eq.~\eqref{eq:CartialFractionsLifted} by a
Gr\"{o}bner basis of the constraint polynomials 
$\{ C_j \}_{j \in A_{q_i}}$.\footnote{It is important to note that Gr\"obner 
basis division works over any
field. Hence, the algorithm can be be applied before or after rational reconstruction.}
After the division, the original polynomial is rewritten as a linear combination 
of the $C_j$ and a unique remainder, which is a polynomial in the $Q_j$ and the $\vec s$.
We then impose
$C_j=0$, leaving this remainder as the result of the 
partial-fraction decomposition. Finally, 
we replace the $Q_j$ by $1/W_j$ and recover an expression of the form 
of eq.~\eqref{eq:LeinartasDecomposition}. We note that
the Gr\"{o}bner basis of the ideal $\{ C_j \}_{j \in A_{q_i}}$ depends on a 
monomial ordering of the variables $\{ Q_j \}_{j \in A_{q_i}}$ and $\{\vec s\}$, 
which in turn affects the final expression. Effectively, by choosing different
orderings one can specify which 
of the denominator factors $\{ W_j \}_{j \in A_{q_{i}}}$ one 
prefers in the final expression. Once this choice has been made,
the result of the procedure is unique.

Let us now comment on why this simple polynomial division technique achieves a
minimal Le\u{\i}nartas decomposition. First, it is important to recall that
the remainder of the division by a Gr\"{o}bner basis is minimal in the sense that
it cannot be written in terms of a member of the ideal and a new remainder
without that remainder having higher polynomial degree. That is, subject to the
choice of monomial ordering, the remainder has the lowest possible polynomial
degree. With this in mind, it is clear why the division by the constraint
polynomials \eqref{eq:PartialFractionConstraints} will maximally cancel
numerator and denominator: since the second term of the constraint polynomial is
$1$, which is the term with the lowest possible degree (in any monomial orderings),
all possible cancellations between numerator and denominator will occur, as that ensures they
will result in the lowest polynomial degree for the remainder.

Another important question is how this division results in a set of denominator
structures with no algebraic dependencies, see eq.~\eqref{eq:algDep}, and 
how the relations are guaranteed to be resolved in a consistent way,
see the discussion around eq.~\eqref{eq:arbitraryLeinartas}.
The fact that algebraic dependencies are implemented follows from a key property of
division by a Gr\"{o}bner basis, which is that it will also apply non-trivial
relations implied by the original set of relations. As already argued in
\cite{raichev2012leinartas}, algebraic dependencies induce such non-trivial
relations. Indeed, if the set of denominator factors of a given term 
\begin{equation}
\label{eq:den}
\frac{1}{W^{\vec \alpha}} \sim Q^{\vec\alpha} 
\end{equation}
are algebraically dependent, then by
definition there exists a non-zero polynomial $P(w)$ such that
$P\big(\{ W_i\}_{i\in A_\alpha}(\vec s)\big) = 0$, which
is trivially a member of the ideal of constraints. 
We associate the same monomial ordering to the $w$ variables as we do to
the $Q_j$ and write the polynomial $P(w)$ in the form
\begin{equation}
    P(w) = c_{\vec\beta}\, w^{\vec\beta} + 
	\sum_{\vec\gamma > \vec\beta} c_{\vec\gamma}\, w^{\vec\gamma}.
\end{equation}
That is, we distinguish the term of lowest monomial degree $\vec\beta$ 
(we refer to the notation introduced in eq.~\eqref{eq:orderingNotation}). 
Let us now multiply the $Q^{\vec\alpha}$ in eq.~\eqref{eq:den} by this polynomial and 
$Q^{\vec\beta}$. We get
\begin{align}\begin{split}
    0=P\big(\{ W_i(\vec{s}\,)\}_{i\in A_\alpha}\big) Q^{\vec \alpha} Q^{\vec \beta}&= c_{\vec\beta}\,Q^{\vec\alpha} \,Q^{\vec\beta} \,W^{\vec\beta}    + 
  \sum_{\vec\gamma > \vec\beta} c_{\vec\gamma} \,
  Q^{\vec\alpha} Q^{\vec\beta}
  W^{\vec\gamma}(\vec s) \\
    &\sim  c_{\vec\beta}\,Q^{\vec\alpha}     + 
  \sum_{\vec\gamma > \vec\beta} c_{\vec\gamma} \,
    Q^{\theta(\vec\alpha + \vec\beta - \vec\gamma)} 
    W^{\theta(\vec\gamma - \vec\alpha - \vec\beta)}(\vec s)\,,
\end{split}\end{align}
where we introduced the function
$\theta(\vec\alpha)$ which yields a vector with the
negative entries removed. This shows that relations such 
as that of eq.~\eqref{eq:arbitraryLeinartas} are now also in the ideal, 
and we do not have to construct them explicitly. Furthermore,
given that by construction $\vec\gamma > \vec\beta$, we
have  
\begin{equation}
\theta(\vec\alpha + \vec\beta - \vec\gamma) < \vec\alpha\,.
\end{equation}
That is, for each term in the sum over $\gamma$, the $Q$-dependent part of the
monomial is always of lower degree than the original denominator 
structure $Q^{\vec{\alpha}}$.
For any monomial ordering where the $Q_j$ are sorted before the $\vec s$, the term  
$c_{\vec{\beta}}\,Q^{\vec{\alpha}}$ is thus leading and will 
therefore be reduced by the polynomial 
division algorithm, which produces remainders of lower degree.
In summary, the algebraic dependence relations are accounted for in the 
constraint ideal, and will be removed by Gr\"{o}bner basis division. 
If, for instance,
one chooses 
a lexicographic order (with the $Q_j$'s sorted before the $\vec s$ variables), 
this will decrease the power of the $Q_j$ for the lowest $j$, 
at the price of potentially increasing the power of $Q_j$ for higher $j$.
This argument can be applied recursively until all algebraic dependencies have been 
resolved. Because the resolution of the dependencies is based on the polynomial ordering
and implemented by the Gr\"{o}bner basis division, it will be
consistently implemented across all denominator factors of a given $r_i^{\pm}$ and
give a unique Le\u{\i}nartas decomposition.

We close by noting that this procedure is both simple to implement as
well as very efficient, easily handling the complex rational functions
present in the problem.  Furthermore, an important consequence of this
decomposition is that the rational numbers in the expression are
more likely to be small. As discussed at the beginning of this section,
this makes the result in a single finite field of cardinality $\mathcal{O}(2^{31})$ 
sufficient to determine them.

\section{Results}\label{sec:results}

In this section we take a closer look at the analytic results we derived
for a basis set of two-loop five-parton amplitudes. 
We list the checks we have performed and describe the format of the ancillary files
which are used to distribute the results. We also comment on the structure
of the remainders.

For each five-parton process we choose a basis set of helicity amplitudes,
such that all other choices of helicities and color structures
from \cref{eq:ColorDecG,eq:ColorDec2Q,eq:ColorDec4Q} can be obtained
by a combination of parity transformation, charge conjugation, and
permutations of external momenta.
Since we reconstruct two-loop remainders, 
in order to assemble two-loop amplitudes
we also need expressions for the one-loop amplitudes through order
$\epsilon^2$, see eq.~\eqref{eq:catani}. Hence, we 
have computed analytic expressions for all the
corresponding leading-color one-loop amplitudes to all orders in $\epsilon$ in
the HV scheme (extending corresponding results of~\cite{Bern:1993mq,
Kunszt:1994nq, Bern:1994fz}) and checked that they reproduce the numerical
one-loop results presented in~\cite{Abreu:2018jgq} up to $\mathcal{O}(\epsilon^2)$.
The calculation of these amplitudes was also performed by analytically reconstructing
the expressions from numerical data.

In order to validate the expressions of the finite two-loop remainders, we have
performed a number of checks. 
First, to check the correctness of the reconstruction procedure, we
have compared our analytic results to an independent numerical evaluation of
the amplitude on rational phase-space points and found agreement. Second,
we have used the analytic results for one-loop amplitudes and two-loop remainders to 
reproduce all available numerical targets~\cite{Badger:2013gxa,
Badger:2017jhb, Gehrmann:2015bfy, Abreu:2017hqn, Badger:2018gip, 
Abreu:2018jgq, Dunbar:2016aux}.
Finally, the analytic results presented here for five-gluon amplitudes at 
$N_f^0$ agree with the previously computed 
expressions~\cite{Gehrmann:2015bfy,Dunbar:2016aux,Badger:2018enw,Abreu:2018zmy}.

We present our results in the form of ancillary files which can be downloaded
from the source files of this work on the \texttt{arXiv} server. We include the
following files, in \Mathematica{}-readable format:
\begin{itemize}
    \item Analytic expressions for the basis set of one-loop five-parton amplitudes
      expressed in terms of one-loop master integrals, valid to all orders in $\epsilon$;
    \item Expressions for all one-loop master integrals written in terms of pentagon functions through order $\epsilon^2$;
    \item The basis set of two-loop five-parton finite remainders, which are the main result
of this work. For each remainder and at each power of $N_f$, we give a list of independent 
coefficient functions $r_i$, a matrix $M_{ij}$ and a list of pentagon functions $h_j$, consistent with the decomposition of eq.~\eqref{eq:remainderDecomposition};
    \item Scripts that assemble two-loop amplitudes from the remainders. 
    In particular, the \Mathematica{} script \texttt{TwoLoopAmplitudesNumerical.m} uses the 
    analytic expressions to reproduce the numeric values of 
    refs.~\cite{Abreu:2018jgq,Badger:2018gip}, as detailed in the included
    \texttt{README.md} file.
\end{itemize}

Our expressions can be easily adapted to perform numerical integration over
phase space. Indeed, they are very compact (with a compressed size of 1.6 Mb),
and ready for automated algorithms for optimized
evaluation like those included in
FORM~\cite{Kuipers:2012rf,Ruijl:2014spa}.
It is interesting to note that whilst there are around 400 independent
pentagon functions, for all amplitudes the number of independent rational
structures is much lower.
A further simplification of the coefficients
can be obtained by combining different powers of $N_f$ in a decomposition
similar to that of eq.~\eqref{eq:remainderDecomposition}. Indeed, in 
figure~\ref{fig:remainderStructure} we observe an overlap of the spaces 
of independent coefficients functions between distinct $\NF$ contributions for a given 
helicity assignment (these are expected from the cancellations 
present in supersymmetric amplitudes). For the most complicated amplitude, 
$(g^-,g^+, g^-, g^+, g^+ )$, the dimension of the combined set of coefficient
functions is only 95. We have not presented the amplitudes in this way in the 
expressions we provide in order to give easier access to the different powers of $N_f$.
While our expressions are specialized to the Euclidean phase-space
region, our coefficients can be used to extract the
required information to cover all regions of phase space with minimal
work. We leave this to a future publication, where we will explore the
numerical evaluation of the amplitudes in more detail.

\begin{figure}
  \centering
  \includegraphics[width = 1.0\textwidth]{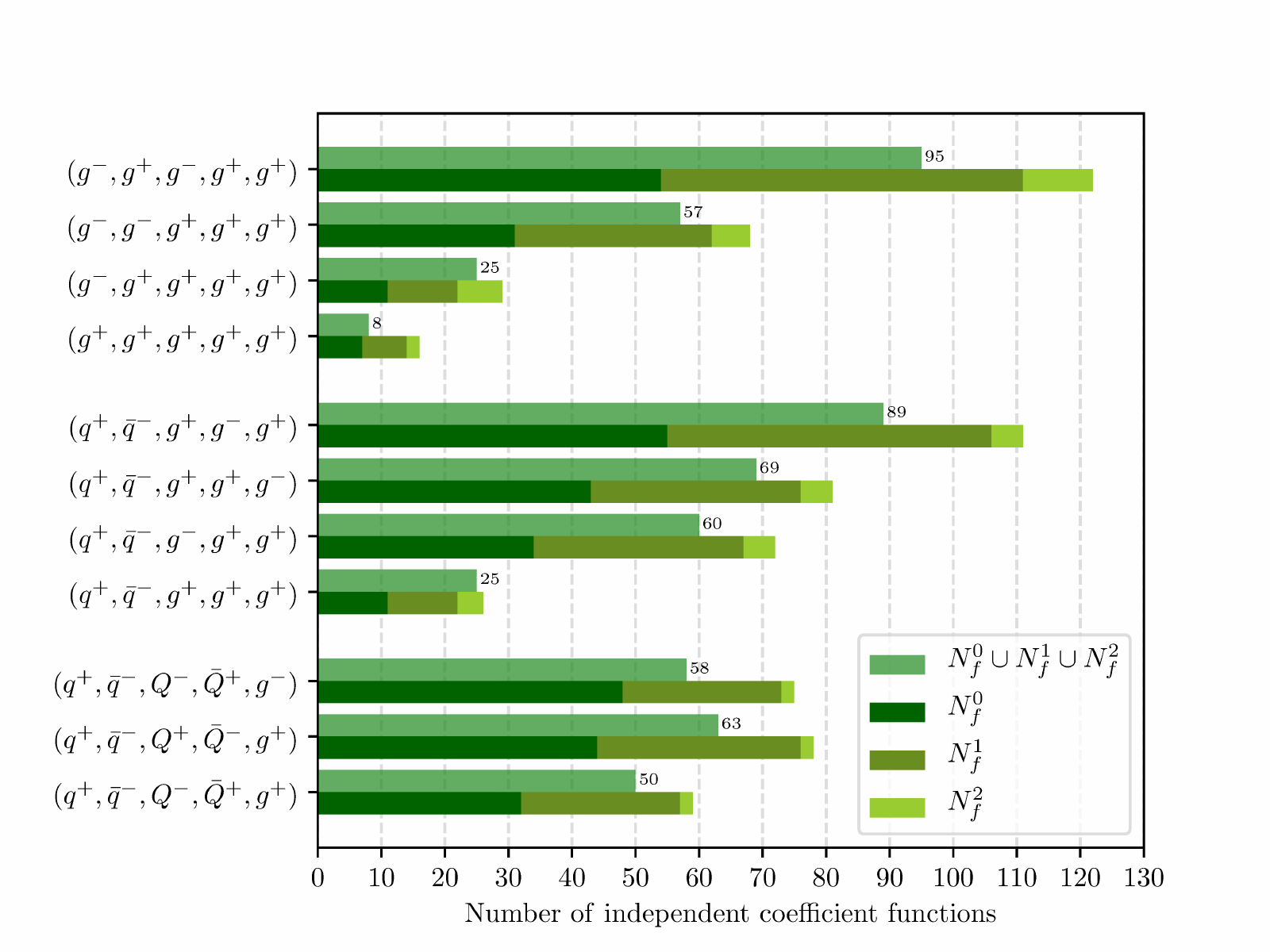}
  \caption{
    The number of linearly-independent coefficient functions $r_i^\pm$ for the five-parton amplitudes.
    The three stacked bars display the case when the three different $N_f$ contributions are considered separately,
    while the bar labeled $N_f^0\cup N_f^1\cup N_f^2$ displays the number of independent functions
    when all $N_f$ contributions are combined.
  }
  \label{fig:remainderStructure}
\end{figure}

We end this section by summarizing the computational resources used
to obtain our results.
The evaluation of the multivariate polynomials $n_i^\pm(\vec s)$ in
eq.~\eqref{eq:DenominatorAnsatz}, which is performed in a single finite field
for all processes, is the most computationally intense.
Every other step of the computation, including the extraction of
denominator factors, the construction of the matrix $M_{ij}$,
\footnote{We note that, for all processes considered, the matrix
$M_{ij}$ was computed with only the information contained in the 
$n_i^\pm(\vec s)$, except for the $N_f^0$ remainder of the processes
$( q^+, {\bar q}^-,Q^-, {\bar Q}^+, g^+ )$ and
$( q^+, {\bar q}^-,g^+, g^-, g^+ )$ where,
in order to rationally reconstruct its
entries, the numerical computation of the remainder over 100 extra phase-space 
points in a second finite field was required.}
performing the partial fractioning of the expressions and rationally reconstructing
the numerical coefficients obtained in the previous step, can be performed
quickly on a modern laptop computer.
The most demanding remainder reconstruction was that of the $N_f^1$ contribution to the
$(g^-,g^+,g^-,g^+,g^+)$ process.
In total, 94,696 phase-space points (coming in parity conjugate pairs,
as discussed in section~\ref{sec:reconstructionInvs}) were
necessary for its reconstruction. On average (considering all contributions to the 
reconstruction procedure) the evaluation time per phase-space point amounts to 4.5 
minutes.\footnote{This timing information is measured on an Intel Xeon E5-2670v3 CPU while 
running the maximum amount of threads it allows.}
The total computational resources required to evaluate all our results are 
relatively modest, and can easily be obtained on a midsize computer cluster.

\FloatBarrier

\section{Conclusion}\label{sec:conclusions}

We have presented the leading-color five-parton two-loop scattering
amplitudes in analytic form for the first time. These are provided
in a set of ancillary files, where we give compact
analytic expressions for five-gluon amplitudes as well as amplitudes
with two and four external quarks, including in all cases the contributions of 
closed light-quark loops. 
These results
have been obtained employing a functional-reconstruction
approach \cite{Peraro:2016wsq} to promote numerical unitarity
\cite{Ita:2015tya,Abreu:2017xsl,Abreu:2017idw} finite-field evaluations of the
amplitude to analytical expressions.

This approach has recently been applied for the planar five-gluon
two-loop amplitudes in ref.~\cite{Abreu:2018zmy} and here we
develop it further in two main directions.  
First, we have improved the handling of fermions in the numerical
approach.  We use dimensional reduction to analytically obtain part of the
dependence on the dimensional-regularization parameter by introducing scalar particles.
This leads to
significant efficiency improvements as compared to a numerical dimensional-reconstruction
approach~\cite{Giele:2008ve,Ellis:2008ir, Boughezal:2011br} as applied earlier
in refs.~\cite{Abreu:2017xsl,Abreu:2017hqn,Abreu:2018jgq}.
Enhancing the dimensional reconstruction by dimensional reduction 
is not a new idea and has been applied to gluon amplitudes in
the original work~\cite{Giele:2008ve}.
Here we apply this improvement~\cite{Anger:2018ove}
starting from the five-point amplitudes including external fermions~\cite{Abreu:2018jgq}.  
Second, we have improved the analytic-reconstruction algorithm from
numerical samples.  We reconstruct directly in terms of Mandelstam
variables as opposed to momentum-twistor variables, which requires
considerably fewer evaluations. Furthermore, we simplify
the reconstruction procedure by focusing on a linearly
independent set of rational functions.
Moreover, we have developed a
multivariate partial-fraction algorithm in order to perform rational
reconstruction with input from only a single finite field of cardinality
$\mathcal{O}(2^{31})$. As a by-product of these techniques, we find compact
analytic forms and an interestingly small set of independent functions.
With these methods, a sample with a size comparable to that required for
numerical Monte-Carlo integration
can be used to produce analytic expressions.

The computational method we present is robust and efficient.
Here, we computed compact analytic expressions for QCD amplitudes depending on
five kinematic scales, which had long been a bottleneck. 
We expect that important scattering amplitudes 
in the Standard Model depending on a higher number of scales 
are within the reach of our approach.
With the recent progress in non-planar master integral
computations~\cite{Abreu:2018aqd,Chicherin:2018old}, it would also be
interesting to explore the amplitudes beyond the leading-color
approximation.

The five-parton amplitudes are an important ingredient for obtaining
precision phenomenology for the three-jet production process at
next-to-next-to-leading order QCD at the LHC in the coming years.
In particular, a complete set of compact analytic expressions, 
such as the ones presented here, will 
be important for an efficient and numerically stable evaluation of the amplitudes.
On a shorter time scale, we believe that our results will be
valuable for exploring the analytic properties of scattering amplitudes
in QCD.

\section*{Acknowledgments}

We thank L.~J.~Dixon, D.~A.~Kosower and M.~Zeng for many useful discussions.
We thank C.~Duhr for the use of his \Mathematica{} package
\texttt{PolyLogTools}.
The work of S.A.~is supported by the Fonds de la
Recherche Scientifique--FNRS, Belgium.
The work of J.D., F.F.C. and V.S. is supported by the Alexander 
von Humboldt Foundation,
in the framework of the Sofja Kovalevskaja Award 2014, 
endowed by the German Federal Ministry of Education and 
Research.
The work of B.P.~is supported by the French Agence Nationale
pour la Recherche, under grant ANR--17--CE31--0001--01.
The authors acknowledge support by the state of Baden-Württemberg through bwHPC
and the German Research Foundation (DFG) through grant no INST 39/963-1 FUGG.
\appendix

\section{Infrared Structure of Two-Loop Five-Parton Amplitudes}
\label{app:remainderDetails}

In this appendix we give more details on our definition of
the remainders we compute in this paper. In 
section \ref{sec:remainders} we stated that divergences of 
renormalized two-loop amplitudes obey a universal structure,
\begin{align}\begin{split}
    \CA_R^{(1)}&={\bf I}^{(1)}_{[n]}(\epsilon)
    \CA_R^{(0)}+\mathcal{O}
    (\epsilon^0)\,,\\
    \CA_R^{(2)}&={\bf I}^{(2)}_{[n]}(\epsilon)A_R^{(0)}+{\bf I}^{(1)}_{[n]}(\epsilon)
    \CA_R^{(1)}+\mathcal{O}(\epsilon^0)\,.
\end{split}\end{align}
The renormalized amplitudes can be written in terms of the
bare amplitudes as
\begin{align}\begin{split}
  \label{eq:twoLoopUnRenorm}
    &\mathcal{A}_R^{(0)}=\mathcal{A}^{(0)}, \\
  & \mathcal{A}_R^{(1)}=S_{\epsilon}^{-1}\mathcal{A}^{(1)}
  -\frac{3}{2\epsilon}\frac{\beta_0}{\NC}
  \mathcal{A}^{(0)}\,,\\
  &\mathcal{A}_R^{(2)}=
  S_{\epsilon}^{-2}\mathcal{A}^{(2)}
  -\frac{5}{2\epsilon}\frac{\beta_0}{\NC}
  S_{\epsilon}^{-1}
  \mathcal{A}^{(1)}
  +\left(\frac{15}{8\epsilon^2}\left(\frac{\beta_0}
  {\NC}\right)^2
  -\frac{3}{2\epsilon}\frac{\beta_1}{\NC^2}\right)
  \mathcal{A}^{
  (0)}\,,
\end{split}\end{align}
with
\begin{equation}
  \beta_0=\frac{\NC}{3} \left( 11 - 2\frac{N_f}{\NC}
  \right),\qquad
  \beta_1=\frac{\NC^2}{3} \left( 17 - \frac{13}{2} 
  \frac{N_f}{\NC} \right).
\end{equation}
For amplitudes in the leading-color approximation the operators
$\mathbf{I}^{(1)}_{[n]}$ and $\mathbf{I}^{(2)}_{[n]}$ are 
diagonal in color space. 
The operator $\mathbf{I}^{(1)}_{[n]}$ is given by
\begin{equation}
  {\bf I}^{(1)}_{[n]}(\epsilon)=
  -\frac{e^{\gamma_E\epsilon}}{\Gamma(1-\epsilon)}
  \sum_{i=1}^n\gamma_{a_i,a_{i+1}}
  \left( -s_{i,i+1}\right)^{-\epsilon}\,,
\end{equation}
with $s_{i,j}=(p_i+p_j)^2$ and the indices defined cyclically.
The index $a_i$ denotes a type of particle with momentum 
$p_i$, i.e., $a_i\in\{g,q,\bar q, Q, \bar Q\}$. 
The symbols $\gamma_{a,b}$ are
symmetric under the exchange of indices, 
$\gamma_{a,b}=\gamma_{b,a}$, and given by:
\begin{align}\begin{split}
  \gamma_{g,g}&=\frac{1}{\epsilon^2}+
  \frac{1}{2\epsilon}
  \frac{\beta_0}{\NC}\,, \qquad
  \gamma_{q,Q}=\gamma_{q,\bar Q}=
  \gamma_{\bar q, Q}=\gamma_{\bar q, \bar Q} 
  =\frac{1}{\epsilon^2}+\frac{3}{2\epsilon}\,,\\
  \gamma_{g,q}&=\gamma_{g,\bar q}=
  \gamma_{g,Q}=\gamma_{g,\bar Q}=
  \frac{\gamma_{g,g}+\gamma_{q,Q}}{2}\,,\qquad
  \gamma_{q,\bar q}=\gamma_{Q,\bar Q}=0\,.
\end{split}\end{align}
The operator~${\bf I}^{(2)}_{[n]}$ is
\begin{equation} \label{eqn:Iop}
    {\bf I}^{(2)}_{[n]}(\epsilon)=
  -\frac{1}{2}{\bf I}^{(1)}_{[n]}(\epsilon)
  {\bf I}^{(1)}_{[n]}(\epsilon)
  -\frac{\beta_0}{\NC\epsilon}{\bf I}^{(1)}_{[n]}(\epsilon) + 
  \frac{e^{-\gamma_E\epsilon}\Gamma(1-2\epsilon)}
  {\Gamma(1-\epsilon)}
  \left(\frac{\beta_0}{\NC\epsilon}+K\right)
  {\bf I}^{(1)}_{[n]}(2\epsilon) + 
  {\bf H}_{[n]}(\epsilon)\,,
\end{equation}
where 
\begin{equation}
K=\frac{67}{9}-\frac{\pi ^2}{3}-\frac{10}{9}\frac{\NF}{\NC}\,,
\end{equation}
and ${\bf H}_{[n]}(\epsilon)$ is a diagonal operator at 
leading color that depends on the number of external quarks 
and gluons in the process,
\begin{align}\begin{split}
  {\bf H}_{[n]}(\epsilon)&=
  \frac{e^{\gamma_E\epsilon}}{\epsilon\Gamma(1-\epsilon)}
  \sum_{i=1}^n\left(
  \delta_{a_i,g}H_g+
  (\delta_{a_i,q}+\delta_{a_i,\bar q}
  +\delta_{a_i,Q}+\delta_{a_i,\bar Q})
  H_q
  \right)\,,
\end{split}\end{align}
with
\begin{align}\begin{split}
  H_g&= \left(\frac{\zeta_3}{2}+\frac{5}{12}+
  \frac{11\pi^2}{144}\right)
  -\left(\frac{\pi^2}{72}+\frac{89}{108}\right)\frac{N_f}{\NC}
  +\frac{5}{27}\left(\frac{N_f}{\NC}\right)^2\,,\\
  H_q&=
  \left(\frac{7\zeta_3}{4}+\frac{409}{864}
  -\frac{11\pi^2}{96}\right)
  +\left(\frac{\pi^2}{48}-\frac{25}{216}\right)\frac{N_f}{\NC}\,.
\end{split}\end{align}

The two-loop bare amplitude ${\mathcal{A}}^{(2)}$
can be obtained from the remainder
defined in eq.~\eqref{eq:remaindeDef} using
\begin{align}\begin{split}\label{eq:ampFromRem}
  {\mathcal{A}}^{(2)}=&\,\mathcal{R}^{(2)}+
  S_\epsilon{\mathcal{A}}^{(1)}
  \left({\bf I}_{[n]}^{(1)}+\frac{5}{2\epsilon}
  \frac{\beta_0}
  {N_c}\right)\\
  &-S_\epsilon^2 {\mathcal{A}}^{(0)}
  \left(
  \frac{15}{8\epsilon^2}
  \left(\frac{\beta_0}{N_c}\right)^2+\frac{3}{2\epsilon}
  \left(\frac{\beta_0}{N_c}{\bf I}_{[n]}^{(1)}-
  \frac{\beta_1}{N_c^2}\right)-{\bf I}_{[n]}^{(2)}
  \right)+\mathcal{O}(\epsilon)\,.
\end{split}\end{align}


\section{Dimensionally Reduced Feynman Rules}
\label{sec:DsFeynRules}
In the table~\ref{tab:Ds-FeynRules} we list the color-ordered Feynman rules for vertices involving the scalar particles introduced in section~\ref{sec:Ds}.
The $(D_s-6)$-dimensional part of these rules can be fully contracted in each Feynman diagram yielding kinematic-independent factors.
\begin{table}[h]
  \centering
  \begin{minipage}[t]{0.4\textwidth}
    \small
  \begin{tabular}{cl}
    $\vcenter{\hbox{\hspace{-2ex}\includegraphics[width=13ex]{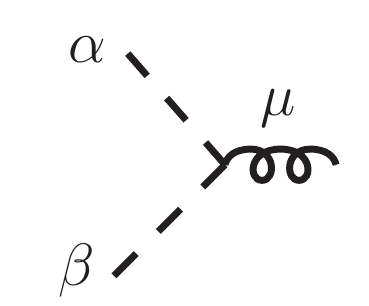}}}$ & $ \frac{i}{\sqrt{2}}(p_2-p_1)^\mu~ g^{\alpha\beta}_{[D_s-6]}$ \\
    $\vcenter{\hbox{\includegraphics[width=15ex]{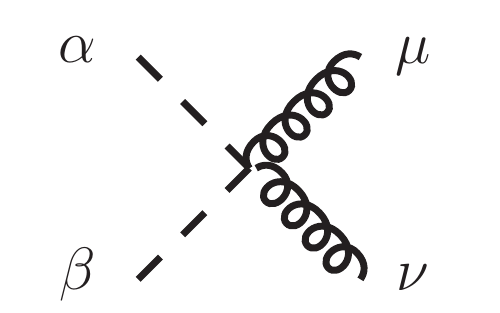}}}$ & $ -\frac{i}{2}~g^{\mu\nu}_{[6]}~g^{\alpha\beta}_{[D_s-6]}$ \\
    $\vcenter{\hbox{\includegraphics[width=15ex]{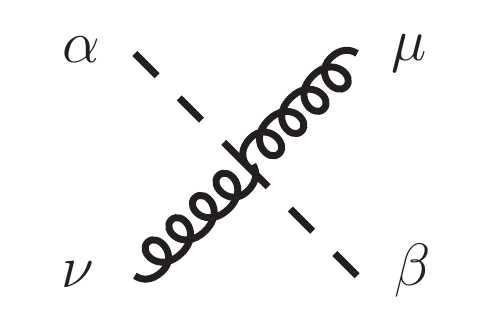}}}$ & $ i~g^{\mu\nu}_{[6]}~g^{\alpha\beta}_{[D_s-6]}$ \\
  \end{tabular}
  \end{minipage}
  \quad
  \begin{minipage}[t]{0.5\textwidth}
    \small
  \begin{tabular}{cl}
    $\vcenter{\hbox{\includegraphics[width=15ex]{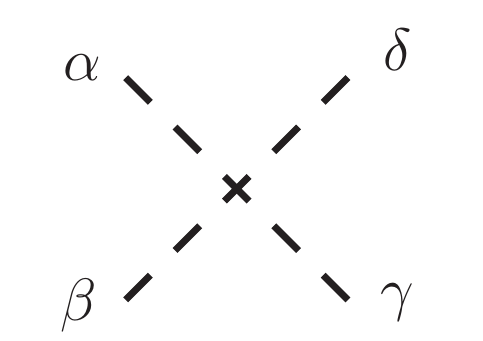}}}$ &
      $
      \begin{aligned}
        i~g^{\alpha\gamma}_{[D_s-6]} g^{\beta\delta}_{[D_s-6]} - ~ & \\
                \frac{i}{2}~(g^{\alpha\beta}_{[D_s-6]} g^{\gamma\delta}_{[D_s-6]} & + g^{\alpha\delta}_{[D_s-6]} g^{\beta\gamma}_{[D_s-6]})
      \end{aligned}
      $ \\
        $\vcenter{\hbox{\includegraphics[width=11ex]{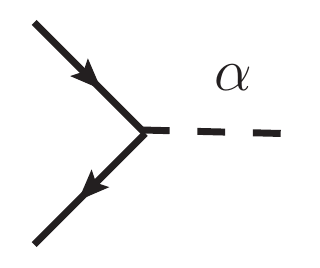}}}$ & $- \frac{i}{\sqrt{2}}~\gamma^\star_{[6]} \gamma^\alpha_{[D_s-6]}$ \\
        $\vcenter{\hbox{\includegraphics[width=11ex]{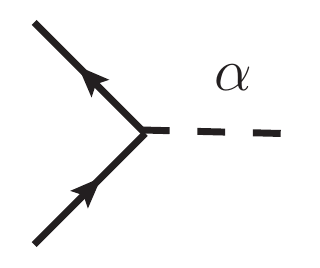}}}$ & $ \frac{i}{\sqrt{2}}~\gamma^\star_{[6]} \gamma^\alpha_{[D_s-6]}$ \\
  \end{tabular}
  \end{minipage}
  \caption{Color-ordered Feynman rules for vertices with scalar particles explicitly introduced by dimensional reduction of gluons.}
  \label{tab:Ds-FeynRules}
\end{table}

\FloatBarrier

\section{Rational Phase-Space Parametrization }\label{sec:twistorVariables}

We give a twistor parametrization~\cite{Hodges:2009hk} used to
rationalize the external on-shell momenta $\{p_i\}_{i=1,5}$ with
$p_i^2=0$. This parametrization yields a momentum point with the
kinematic invariants $(s_{12}, s_{23}, s_{34}, s_{45}, s_{51})$ from the
input variables $\{ s_{23}, s_{45}, s_{51}, x \}$ and $s_{12}=1$ 
as given in the main text in eq.~\eqref{eq:s34Definition}.

To each external momentum $p_i$, we associate the spinors $\lambda_i$
and $\tilde{\lambda}_i$, from which we can compute the associated
momenta through~
\begin{equation}
p_i^\mu = \frac{1}{2} \,\tilde{\lambda}_i^T \sigma^\mu \lambda_i   \, .
\end{equation}
The spinors are then parametrized through the momentum-twistor matrix,
\begin{equation}
\left(
            \begin{matrix}
\lambda_1 &  \lambda_2 &   \lambda_3 &   \lambda_4 &  
\lambda_5 \\   
\mu_1 &  
\mu_2 &  
\mu_3 &  
\mu_4 &  
\mu_5 
\end{matrix}
\right)
=
\left(
            \begin{matrix}
            1 & 0 & 1 & 1 \!+\! \frac{1}{x}  &  1 \!+\! \frac{1}{x} \!+\! \frac{x-s_{23}+s_{45}}{x s_{51}}\\
            0 & 1 & 1 & 1                    &  1\\
            0 & 0 & 0 & \frac{s_{23}}{x}     &  1\\
            0 & 0 & 1 & 1                    & 1 \!-\! \frac{s_{45}}{s_{23}}
            \end{matrix}
            \right).
\label{eq:TwistorParametrization}
\end{equation}
Conjugate spinors $\tilde{\lambda}_i$ can then be obtained through
\begin{equation}
\tilde{\lambda}_i = \frac{\langle i, i+1 \rangle \mu_{i-1} + \langle i+1, i-1\rangle \mu_{i} + \langle i-1, i \rangle \mu_{i+1}}{\langle i, i+1\rangle \langle i-1, i\rangle}\, ,
\end{equation}
where $\langle i, j \rangle = \det(\{\lambda_{i},\lambda_{j}\})$ 
is the usual spinor-bracket computed by taking the
determinant of the associated sub-matrix of
\eqref{eq:TwistorParametrization}. The Mandelstam invariants $s_{ij}$ and $\trFive$ required in the main text are given in terms of spinors by,
\begin{equation}
		s_{ij} = \langle i,j \rangle \, [j,i]\,,\quad 
		\trFive=[1,2]\langle 2,3\rangle [3,4] \langle 4,1\rangle 
	-\langle 1,2\rangle [2,3] \langle 3,4\rangle [4,1]\,,
\end{equation}
using 
$[ i ,j ] = \det(\{ \tilde{\lambda}_{j},\tilde{\lambda}_{i}\})$.
For further details, see  e.g. section 2 of ref.~\cite{Bourjaily:2010wh}.

\bibliography{main.bib}

\end{document}